\PassOptionsToPackage{pdfpagelabels=false}{hyperref}

\documentclass[preprint,authoryear,5p]{elsarticle}

\usepackage{newtxtext,newtxmath}
\usepackage[T1]{fontenc}
\usepackage{newtxtext,newtxmath}
\usepackage{ae,aecompl}
\usepackage{placeins}
\usepackage{graphicx}	% Including figure files
\usepackage{amsmath}	% Advanced maths commands
\usepackage{amssymb}	% Extra maths symbols
\usepackage{hyperref}
\usepackage{gensymb}
\usepackage{natbib} % biblio
\usepackage{lineno} % for submission to icarus
\usepackage{url,wasysym}
\usepackage{subcaption}
\usepackage{longtable}

\title{Ricochets on Asteroids: Experimental study of low velocity
grazing impacts into granular media}

\author[a1]{Esteban Wright\corref{cor1}}
\ead{ewrig15@ur.rochester.edu}
\author[a1]{Alice C. Quillen}
\ead{alice.quillen@rochester.edu}
\author[a1]{Juliana South}
\ead{jsouth@rochester.edu}
\author[a2]{Randal C. Nelson}
\ead{nelson@cs.rochester.edu}
\author[a3]{Paul S\'anchez}
\ead{diego.sanchez-lana@colorado.edu}
\author[a1]{John Siu}
\ead{jsiu3@u.rochester.edu}
\author[a4]{Hesam Askari}
\ead{askari@rochester.edu}
\author[a1,a5]{Miki Nakajima}
\ead{mnakajima@rochester.edu}
\author[a6,a7]{Stephen R. Schwartz}
\ead{srs51@email.arizona.edu}
\address[a1]{Department of Physics and Astronomy, University of Rochester, Rochester, NY 14627, USA}
\address[a2]{Dept. of Computer Science, University of Rochester, Rochester, NY, 14627, USA}
\address[a3]{Colorado Center for Astrodynamics Research, The University of Colorado Boulder, 3775 Discovery Drive,
429 UCB - CCAR,
Boulder, CO 80303, USA}
\address[a4]{Department of Mechanical Engineering, University of Rochester, Rochester, NY 14627, USA}
\address[a5]{Department of Earth and Environmental Sciences, University of Rochester, Rochester, NY 14627, USA}
\address[a6]{Lunar and Planetary Lab, University of Arizona, Tucson, AZ, USA}
\address[a7]{Laboratoire Lagrange, Universit\'e C\^ote d'Azur, Observatoire de la C\^ote d'Azur, CNRS, C.S. 34229, 06304 Nice Cedex 4, France}

\cortext[cor1]{Corresponding author}

\begin{document}

\begin{abstract}

Spin off events and impacts can eject  boulders from an asteroid surface and rubble pile asteroids can accumulate from debris following a collision between large asteroids.
These processes produce a population of gravitational bound objects in orbit that can impact an asteroid surface at low velocity and with a distribution of impact angles. 
We present laboratory experiments of low velocity spherical projectiles into a fine granular medium, sand.   
We delineate velocity and impact angles giving ricochets,  those giving projectiles that roll-out from the impact crater and those that stop within their impact crater. 
With high speed camera images and fluorescent markers on the projectiles we track spin and projectile trajectories during impact.  We find that the projectile only reaches a rolling without slipping condition well after the marble has reached
peak penetration depth.
The required friction coefficient during the penetration phase of impact is 4-5  times lower than that of the sand suggesting that the sand is fluidized near the projectile surface during penetration.
We find that the critical grazing impact critical angle dividing ricochets from roll-outs, increases with increasing
impact velocity.  The critical angles for ricochet and for roll-out as a function of velocity can be matched by an empirical  model during the rebound phase that balances a lift force against gravity.
We estimate constraints on projectile radius, velocity and impact angle that would allow projectiles on asteroids to ricochet or roll away from impact, finally coming to rest  distant from their initial impact sites. 
%We compare our laboratory results to obsereved tracks on Eros by \cite{durda12}.

\end{abstract}

\begin{keyword}
%Keywords:
\end{keyword}

\maketitle
%\linenumbers %  turn this on for submission 

\section{Introduction}

Impact crater ejecta curtains (e.g., \citealt{asphaug93,thomas01,durda12}), seismicity associated with impacts (e.g. \citealt{wright19}), and mass loss associated with spin-off events (e.g., \citealt{holsapple10,hirabayashi15}) are processes that would eject a population of rocks 
and boulders from an asteroid surface.  Eventually, this material would be ejected from the asteroid's vicinity or it would return to hit the asteroid surface.    A rubble pile asteroid can be formed following a disruptive collision of two large bodies,
including a phase of re-accumulation from previously disrupted but gravitationally bound material \citep{michel13,walsh18,walsh19}. 
Late stages of re-accumulation involves low velocity impacts onto the asteroid surface.
Impacts with objects in the main asteroid belt have a mean relative
velocity of $\sim 5$ km/s \citep{bottke94}.   
In comparison,  a gravitational bound object would impact the asteroid surface at much lower velocity,
less than the escape velocity which is $\sim 20$ cm/s for an 500 m diameter object such as Asteroid 101995 Bennu \citep{scheeres19}. These would be low velocity impacts into surface regolith or rubble at low gravitational acceleration and  would include encounters at low or
at grazing incidence angle because the projectiles were previously in orbit.  Due to its orbital angular momentum,
this would also be true of most material that was thrown off the asteroid during spin-off events.
Boulders ejected during crater formation can also contribute to the population of objects returning to hit the surface at low velocity and grazing incidence angle \citep{durda12}.
%$\sim 10^{-4} $ to $10^{-5} $ g for Asteroid 162173  Ryugu and Bennu, with the range dependent on latitude due to asteroid spin and centripetal acceleration. 

Experiments of low velocity and normal impacts into granular media at low surface gravity from drop towers include those by 
\citet{goldman08,Sunday_2016,murdoch17} and in aircraft those by \citet{brisset18}.
%The range of gravity regimes for low velocity impact experiments has recently been  extended by \citet{Sunday_2016},  who performed drop-tower experiments with an effective gravity going down to 0.01g. 
These experiments studied normal impacts, those perpendicular to the impact surface, and so did not study the sensitivity of the projectile deceleration profile and penetration depth to impact angle. The deceleration of the projectile is due to interaction within the impacted surface.
However debris that is orbiting an asteroid, that later hits the asteroid surface, would be unlikely to only have high or nearly normal impact angles. Grazing impacts of spherical projectiles on sand or water are more likely to bounce or ricochet  (e.g., \citealt{birkhoff44,soliman76,daneshi77,bai81}).  Simulations by \citet{maurel18} show that this is also true for non-spherical, low velocity projectiles, such as a lander, in low surface gravity.

\begin{figure}
%\centering
\includegraphics[width=3.5in]{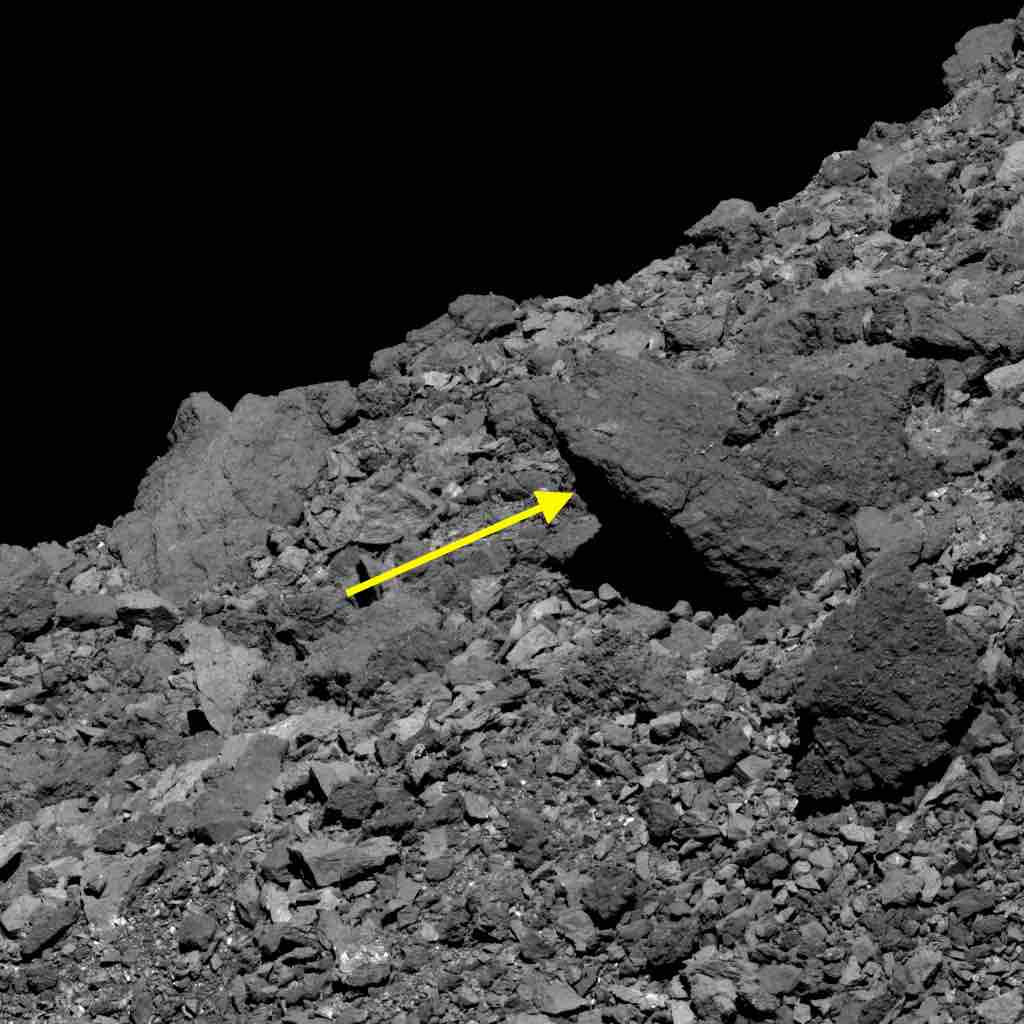}
\caption{A perched boulder is 47 ft (14.3 m) long on Bennu. From 
\url{https://www.asteroidmission.org/20190405-shelf/}
This image was taken by the PolyCam camera on NASA's OSIRIS-REx spacecraft on April 5 2019 
from a distance of 2.8 km. The field of view is 40.5 m.
\label{fig:bennu}}
\end{figure}

Images of asteroids 101995 Bennu and 162173 Ryugu show large (10--50 m) boulders that look as if they were perched on the surface \citep{sugita19,walsh19}. See Figure \ref{fig:bennu} for a boulder that might unstable if perturbed by large amplitude vibrations.
Possible explanations for boulders on the surface of a rubble pile asteroids include the Brazil nut effect (e.g., \citealt{matsumura14}), which drives previous buried boulders to the surface,  and  boulder stranding that occurs during landing of ejecta that is launched by an impact generated pressure pulse \citep{wright19}. Some of the surface boulders on Bennu are so large they probably instead landed on the surface during accumulation following a disruptive large collision of larger asteroids \citep{walsh19}.  A population of large boulders ($\sim 30$ m diameter) littering asteroid 433 Eros' equatorial region,  is attributed to ejecta from the impact event that formed
the Shoemaker crater \citep{thomas01}.

A low velocity boulder projectile with a shallow or grazing impact angle might bounce off the surface, landing distant from the site of first impact and from any depression in the surface resulting from material ejected during its first impact. Boulders that were ejected by an impact can
leave oblong tracks or secondary craters
where they bounce off the surface.   Examples of such tracks and associated 2, 25 and 40 m diameter boulders on asteroid 433 Eros suggest that the boulders were emplaced at the termination of their trajectories \citep{sullivan02,durda12}.
% A few good examples of boulders with tracks (Fig. 1) can be found on Eros (Sullivan et al. 2002).
We consider ricochets of low velocity impactors as a way to account for protruding boulders on the surfaces of asteroids such as Eros, Bennu and Ryugu. While Bennu and Ryugu might lack surface regions comprised primarily of fine grained material, other asteroids such as 433 Eros \citep{veverka01,cheng02},  25143 Itokawa \citep{miyamoto07} and the moon exhibit smooth regions covered in regolith. The dynamics of low velocity grazing impacts into granular media is also relevant to interpretation of the surfaces of these bodies and for the design of landers that may be going to them. For a review of granular media in solar system bodies see \citet{hestroffer19}.

Deployed from the European Space Agency's 
Rosetta spacecraft, the Philae lander's
anchoring harpoons failed to fire when approaching comet 67P/Churyumov-Gerasimenko.      
The lander  rebounded twice from the comet surface prior to coming to rest  roughly 1 km away from the intended landing site \citep{biele15}.   The first touchdown was at a relative velocity 
of 1\ m/s (and near the local escape velocity) and about $12^\circ$
from normal.  The normal velocity component was damped and the outgoing velocity was about 1/3 of
the incoming relative velocity.  
The lander itself contains a damping element which was depressed during
this touchdown.  Interpretation of the vertical acceleration profile suggest that the lander 
hit a granular surface  with a compressive strength of order a few kPa \citep{biele15}.
%The first touchdown site, Agilkia, appears to have a granular soft surface (with a compressive strength of 1 kilopascal) at least ~20 cm thick, possibly on top of a more rigid layer. The final landing site, Abydos, has a hard surface.
%\citep{tsuda13} for Rosetta mission

Hyabusa2 is a sample return mission to asteroid Ryugu that contains a 10kg lander called MASCOT (Mobile ASteroid SCOuT;  \citealt{ho16}).
The soft sphere rubble pile simulations by \citet{maurel18} showed that  simulated low velocity impacts (19 cm/s) of MASCOT onto Ryugu's surface  with lower (closer to grazing) impact angles were more likely to bounce and had higher effective coefficients of restitution than normal impacts. The Philae and MASCOT landers illustrate that the understanding of low velocity surface-lander interactions is critical for missions with lander components and influences lander deployment and sample return strategies.
% \citet{maurel18} use an angle of approach to characterize simulated impact angles  and these are measured  as between the impact trajectory and the surface normal.

Phenomenological models have been proposed to account for experimental measurements of penetration depth of spherical projectiles impacting granular materials in a gravitational field at normal incidence (e.g., \citealt{uehara03,tsimring05,ambroso05,katsuragi07,goldman08,katsuragi13,murdoch17}).
For normal impacts (e.g., \citealt{uehara03,tsimring05,ambroso05,goldman08,katsuragi13,murdoch17}) and recent oblique impact experiments \citep{bester19}.
Most of these experimental studies have modeled the granular medium with an empirical force law that includes a hydrodynamic drag term proportional to the square of the velocity, which dominates at higher velocity and so at deeper penetration, and a term which accounts for a depth dependent static resistance force, which dominates at lower speeds and shallow penetration depths.
However, a higher velocity projectile at grazing incidence may not penetrate deeply and would still feel a hydrodynamic-like drag.
Horizontal motion in granular media can cause lift \citep{ding11,potiguar13}.
Thus the empirical models primarily developed for normal impacts need to be modified in order to account for impacts at grazing angles. 

Phenomenological models for ricochet in sand or water \citep{birkhoff44,johnson75,daneshi77,soliman76,bai81} predict a velocity dependent critical angle for ricochet.  However, these early models assume low angles of impact and nearly horizontal and constant velocity during the encounter between projectile and substrate and neglect spin.  The lift was based on a hydrodynamic pressure exerted on the submerged projectile surface and may not be  consistent with  more recent empirical force models or experimental measurements. 
%and measurements of stopping depths and times from normal impact experiments (e.g., \citealt{tsimring05,katsuragi07,goldman08,murdoch17}).
Models consistent with a broader range of phenomena,  including ricochets and oblique impacts, would be helpful for understanding processes taking place in low gravity environments.

\section{Laboratory Experiments of Projectiles Impacting Similar Density Granular Material}
\label{sec:exp}

We carry out experiments of spherical projectiles into sand.
We restrict the density of our projectiles to similar materials and densities as the solids in our substrate material so as to be similar to natural low velocity impacts on asteroids. Prior to this study we had explored non-spherical projectiles (pebbles) into coarser substrates (gravel), however we found that experiments were often not easily reproducible.  This was likely due to the complex projectile shapes, their spin and phase of impact with respect to rotation, and irregularities in the substrate.  We have reduced the degrees of freedom so as to try to understand simpler systems as a first approximation. The projectiles we discuss here are spheres and 
the granular substrate is comprised of particles that are much smaller than the projectile.

Our granular substrate is dry playground sand. As  irregularities in a fine medium are less likely to affect the projectile trajectory, and craters are easier to see in a fine medium, we chose fine sand to facilitate measurement of projectile stopping times,% crater  sizes,  
crater depths and morphology.  
The sand substrate had a depth of 6.3 cm and was passed through a sieve, giving only particles less than 0.5 mm in diameter. Density of our sand is $\rho_s = 1.6$ g/cm$^3$.  
%The average density of quartz (solid silicon dioxide) is $\sim 2.6$ g/cc, so the porosity of the sand is approximately 0.5.  
The porosity of our sand was 0.4. We measured the porosity by filling a volume of sand with water until the voids were filled and taking the ratio of the volumes.
Prior to each impact the substrate is raked and then scraped flat.  It is not pressed or compacted. The rake consists of a linear row of 2 cm long nails. The separation between each pair of neighboring nails is 1 cm. 
The rake and a close up view of the sand tray is shown in Figure \ref{fig:setup_close}.
The low packing fraction of the sand and shallow penetration regime of our impacts suggests that interstitial air would not have a strong affect the impact dynamics (see \citealt{royer11}).

We use a small spherical projectile. The glass marble projectile we used for most experiments has a mass of 5.57 g, a diameter of 16.15 mm and a density of 2.5 g cm$^{-3}$.  This density is within the scatter of different types of quartz and glasses.

It is useful to define
a dimensionless number
that will scale our laboratory experiments on Earth to an asteroid setting.
We chose a dimensionless number known as the Froude number
\begin{equation}
 Fr \equiv \frac{v}{\sqrt{gR_p}}   \label{eqn:Fr}
\end{equation}
with $R_p$ the projectile radius.
The Froude number is related to the dimensionless number, $\pi_2$, used to describe impact crater scaling relations \citep{housen03}; $\pi_2 = 3.22 g R_p/v^2 = 3.22 Fr^{-2}$.
A small size in the lab is convenient as a  $R_p \sim 1$ cm radius projectile at an impact speed of $v_{impact} = 3$ m/s has dimensionless Froude number $Fr \equiv v_{impact}/\sqrt{g R_p} \sim 10$ that is similar to that of a meter radius projectile at an impact speed similar to the escape velocity $\sim 20 $ cm/s on a small asteroid like Bennu with effective gravity $\sim 10^{-4}$ g.

\begin{figure}
%\centering
\includegraphics[width=3.4in]{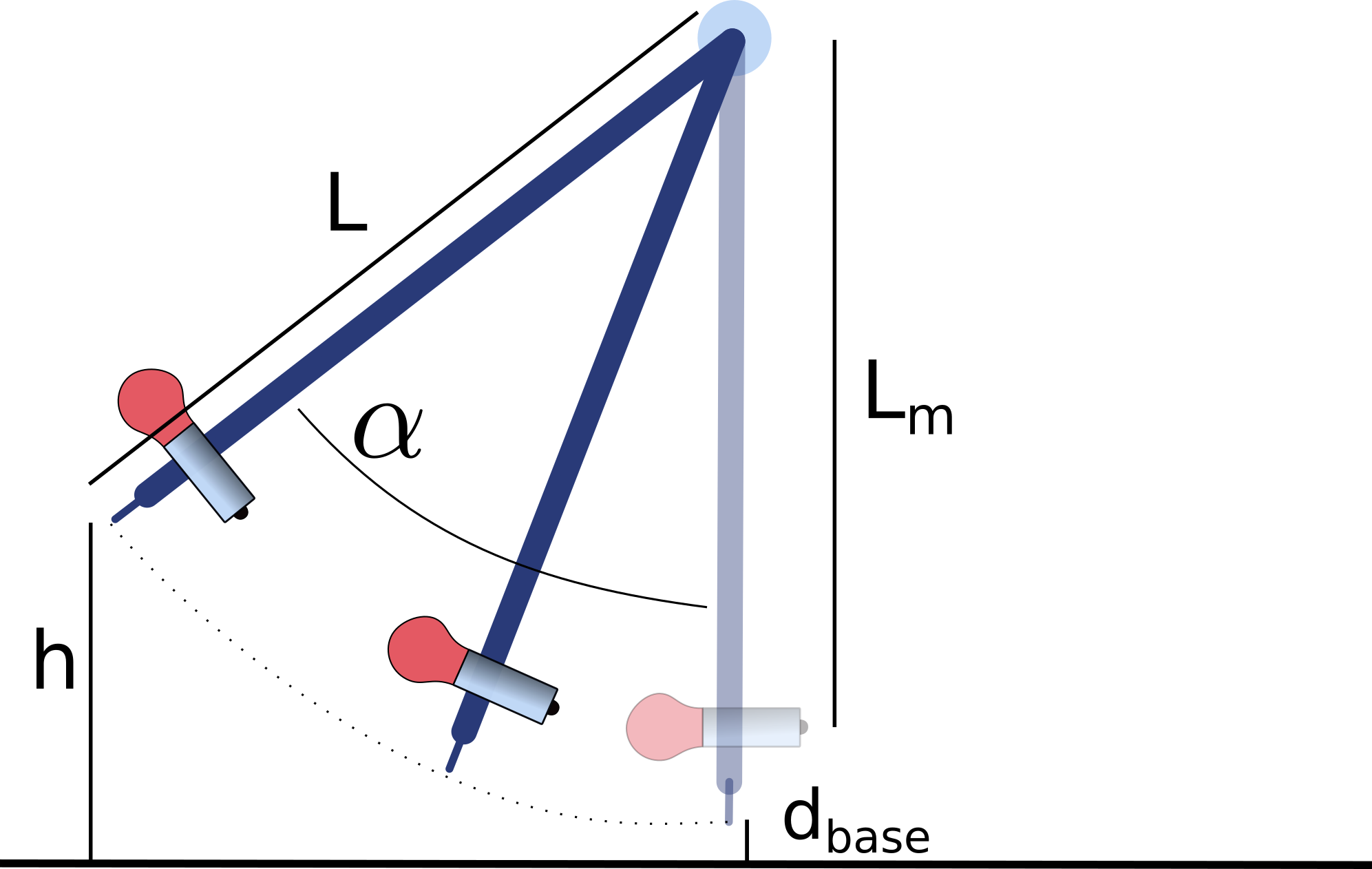}

\vspace{0.5cm}

\includegraphics[width=3.4in]{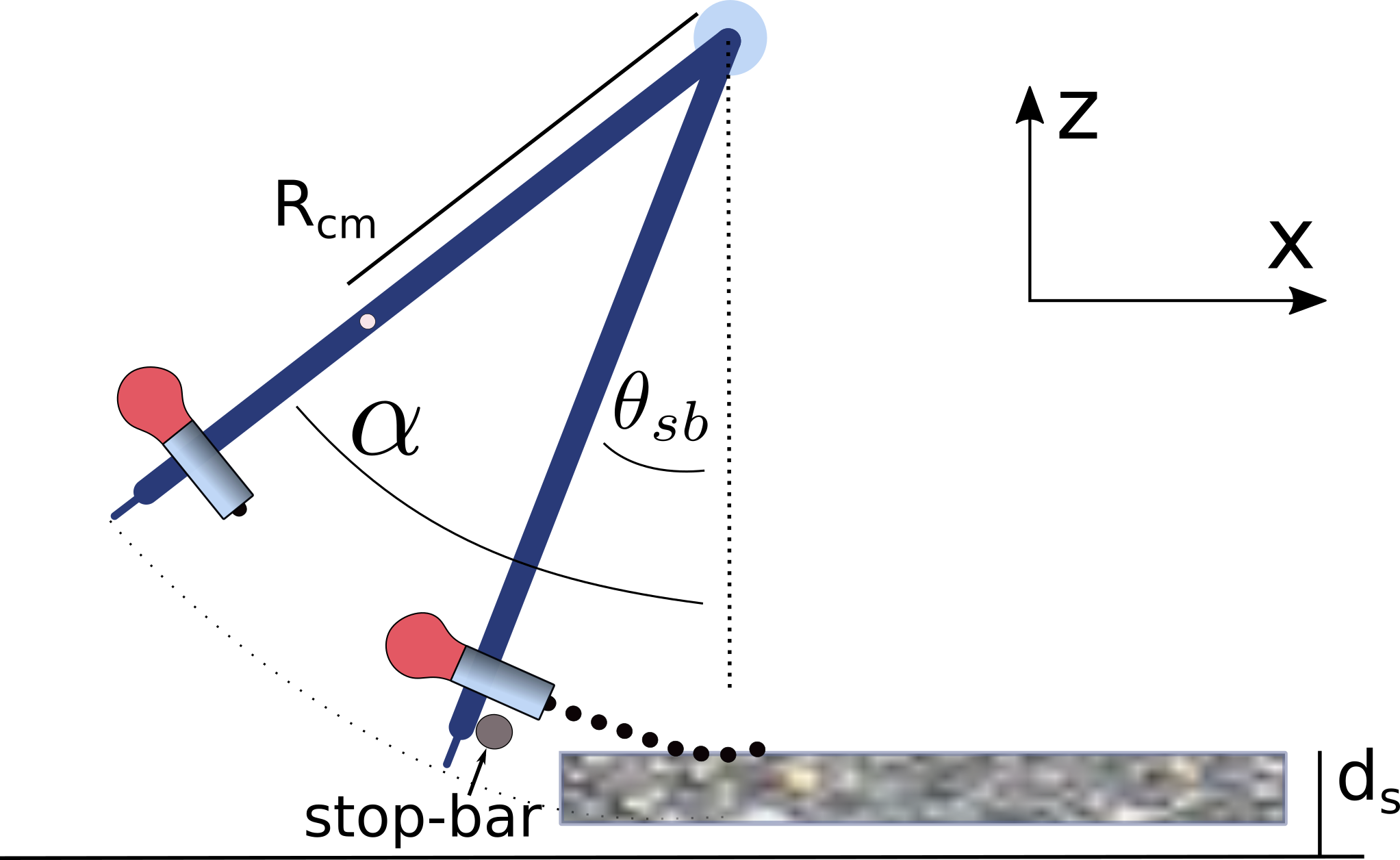}
\caption{Side-view illustrations of our pendulum based marble launcher. The pendulum is dropped from a height $h$, setting the velocity of impact.  The pendulum swing is stopped by a stop-bar at an angle $\theta_{sb}$ that sets the angle of impact. The marble is ejected from its holder when the pendulum is stopped and then hits the sand. The red bulb is used to apply a weak vacuum that holds the marble in place  until the pendulum hits the stop-bar.  The lengths and angles shown here are used to estimate the velocity and angle of the marble impact on the sand. See Table \ref{tab:pend} for a list of quantities and Table \ref{tab:nomen} for nomenclature.
\label{fig:pend}
}
\end{figure}

\begin{figure}
%\centering
\includegraphics[width=3.4in]{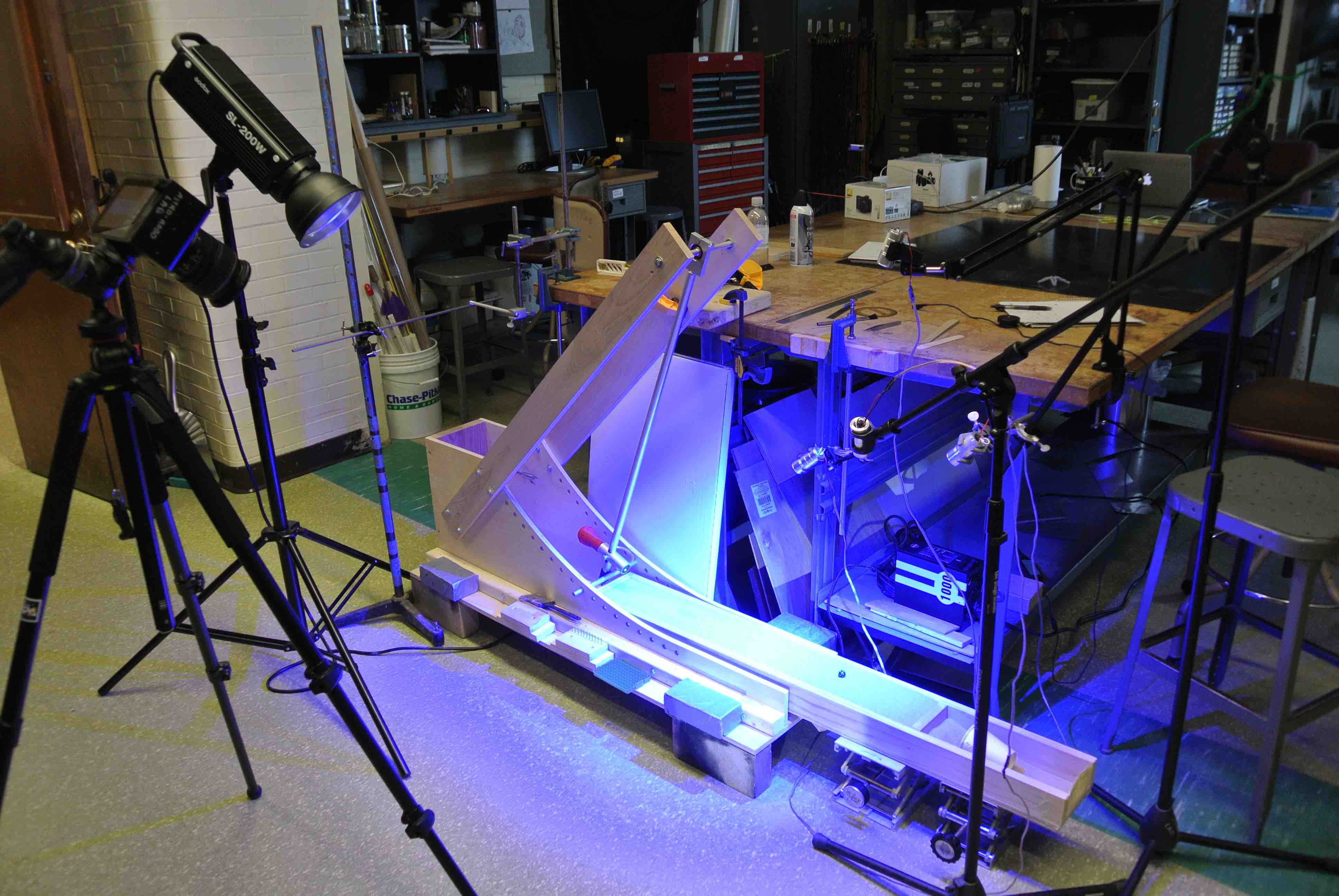}
\caption{Photograph of the experimental setup. The high speed camera and white light are on the left, whereas the blue LEDs are on the right.  The sand box is just above the floor with inside dimensions of 87.5 cm long,  11.5 cm  wide and 6.3 cm deep.
\label{fig:setup}
}
\end{figure}

\begin{figure}
%\centering
\includegraphics[width=3.5in]{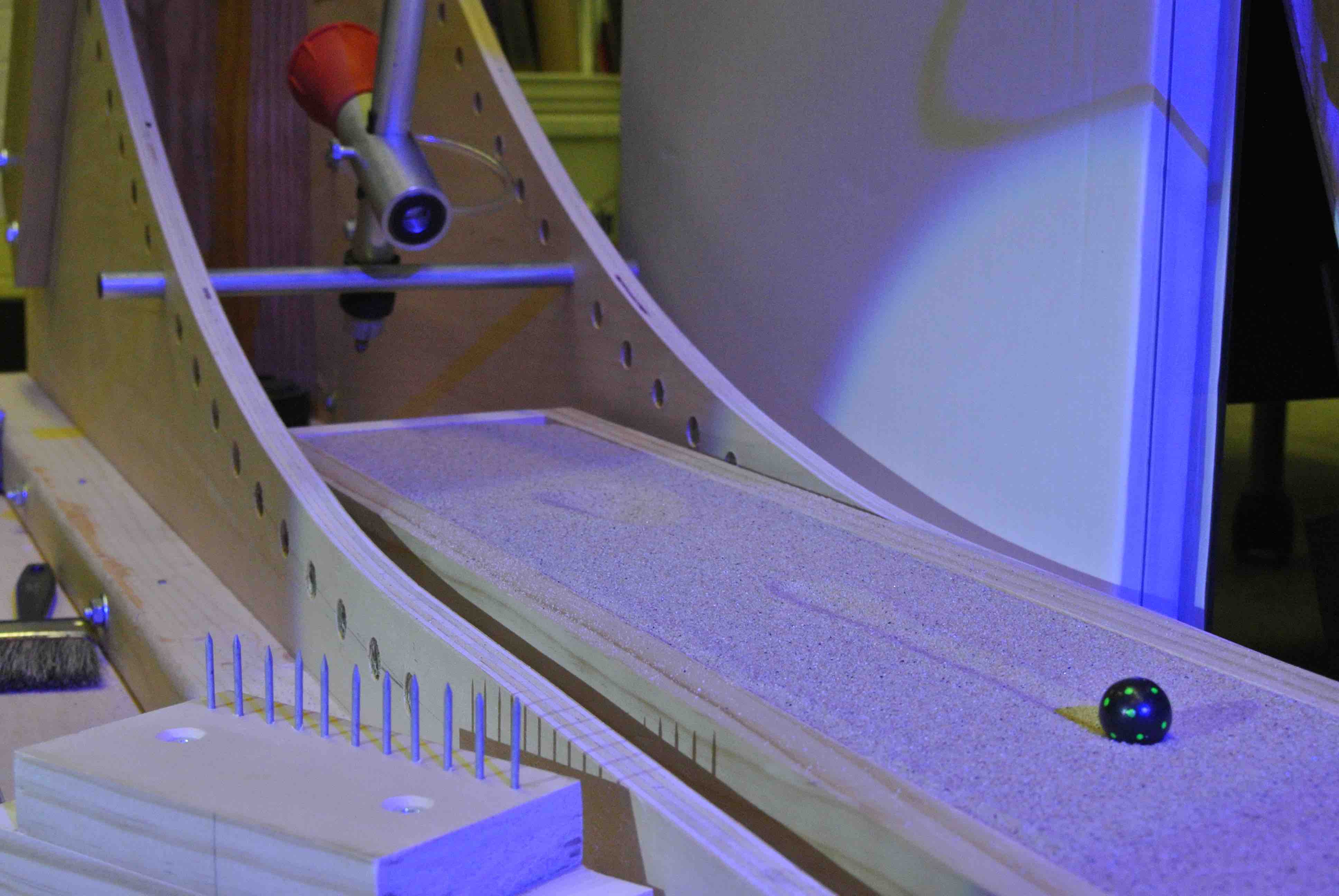}
\caption{Close up view of the sand tray and pendulum launcher. The marble in the foreground is 16 mm in diameter. Prior to each experiment we raked the sand flat.  The rake we used is shown on the left and has nails that  penetrate to a depth of 2 cm. 
\label{fig:setup_close}}
\end{figure}

\begin{figure}
%\centering
\includegraphics[width=3.4in]{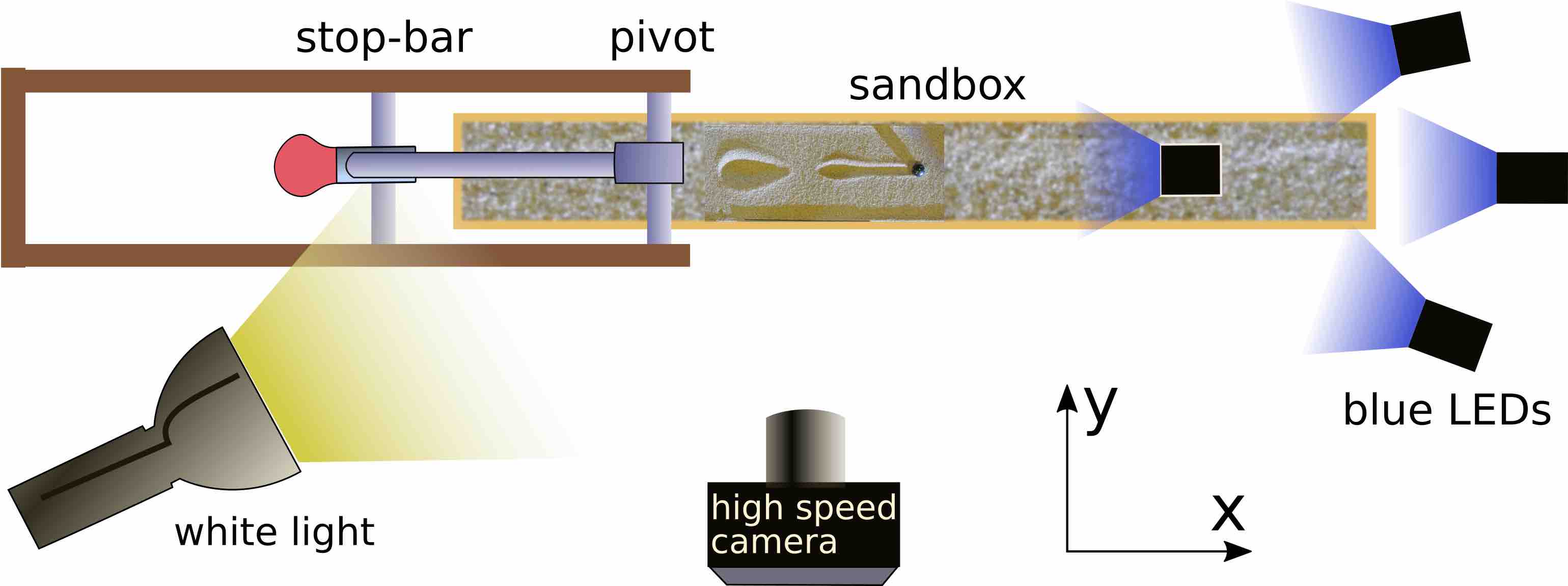}
\caption{A top view of the pendulum marble launcher showing the lighting and high speed camera viewing position.
\label{fig:ric_lighting}
}
\end{figure}

We desire a way to launch projectiles that minimizes initial spin, allows us to adjust impact velocity and impact angle and gives reproducible craters and outcomes (ricochet or not).   After trying a rail gun, we settled on a pendulum launcher because it gave us low projectile spin and more repeatable impact velocities and angles. The pendulum is raised to a set height and then drops due to gravity until it hits a horizontal stop-bar.  The location of the bar
that stops the pendulum as it swings down sets the projectile impact angle.
Our pendulum setup is illustrated in Figure \ref{fig:pend}.  A photograph of the experimental setup is shown in Figure \ref{fig:setup} and a close up view in Figure \ref{fig:setup_close}.  An illustration of the sandbox, camera and lighting as viewed from above is shown in Figure \ref{fig:ric_lighting}.

Prior to letting the pendulum drop, the marble is held up against a thin rubber washer.
A red turkey baster handle is used to apply light suction to the marble to hold it in place.
%While the pendulum swings, the marble is held in place with this  suction.
A light tap on the pendulum  is enough to break the suction and
eject the marble. 
Because the required tap is light, the marble suffers only a small reduction in velocity during ejection.
The marble is ejected with some backspin (angular rotation rate $\sim -30$  rad  s$^{-1}$),  that is usually well  below the size of the horizontal velocity component divided by marble radius  ($\sim $ 100 to 300 rad s$^{-1}$).
Backspin is larger than expected from the pendulum swing which would be $v_{eject}/L_m \sim 5$ rad s$^{-1}$.
$v_{eject}$ is the velocity of the marble when ejected from launcher, and $L_m$ is the radial distance from the pivot to the ejection point.
The backspin must be caused by uneven friction or suction when the marble exits its holder.
Measurements for the marble spin are discussed in more detail in section \ref{sec:data}.

Craters in our sandbox were increasingly reproducible after mechanical vibrations were reduced. The pendulum rod is a hollow aluminum pipe, replacing a narrower and heavier steel rod that flexed upon impact.  The aluminum rod still flexes some on impact and this might be the cause of the ejected marble's backspin. The pendulum pivot is clamped to a lab table to prevent bounces and vibration during the impact between pendulum and horizontal stop-bar.
We adjusted the angle of the marble holder so that the marble is ejected in the same plane of the pendulum.
The connection between marble holder and pendulum rod was shimmed and tightened so that it did not rotate.
We added super glue to the the thread at the top of the pendulum rod to prevent it from turning. 
Prior to each impact experiment, we checked that the sand tray is leveled in directions parallel and perpendicular to the marble launcher using a bubble level.

The pendulum itself is $L=94.8$ cm long, but the radial distance from pivot to marble ejection point is $L_m = 84.3$ cm long.  
Due to its extended mass distribution, the pendulum is a compound pendulum.
We measured the moment of inertia of the pendulum from its period of small oscillations ($T=1.78$ s) and its center of mass radius from the pivot, $R_{cm} = 50.5$ cm using the relation
\begin{equation}
\frac{I_{pend}}{M_{pend}} = \frac {T^2}{(2\pi)^2} g R_{cm}
\end{equation}
where $M_{pend}$ and $I_{pend}$ are the mass and moment of inertia (about the pivot point) of the pendulum. 
Measurements of the pendulum are listed in Table \ref{tab:pend}.
The inside dimensions of the sandbox are 87.5 cm long,  11.5 cm  wide and 6.3 cm deep.
%These are inside dimensions of the box and give the total volume of sand.

% pendulum lighting figure

% crater still images
\begin{figure}
\centering
\begin{subfigure}[b]{0.95\linewidth}
   \centering
   \includegraphics[width=1\linewidth]{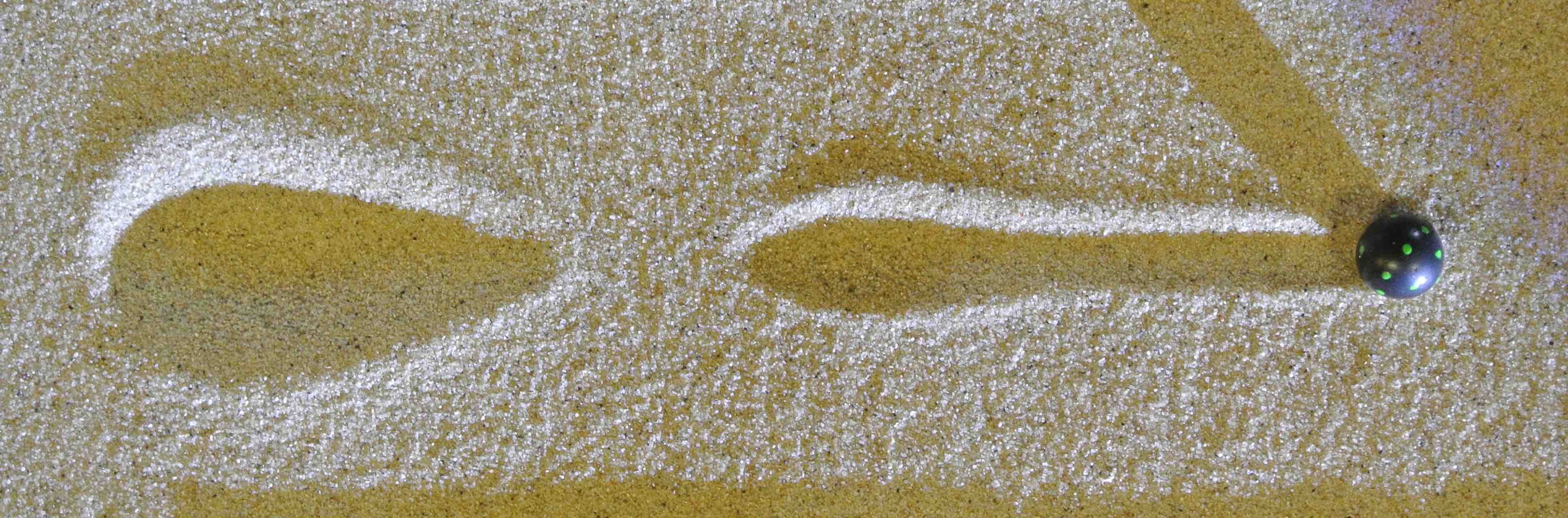}
   \caption{Ricochet}
   %{Impact crater of experiment 30 with a 1.6 cm diameter marble, an example of a ricochet event. The diameter of the initial crater is 8.95 cm wide by 5.15 cm tall and 0.95 cm deep. The gap between the initial crater and the marble trail is 3 cm.}
   \label{fig:crater_ricochet} 
\end{subfigure}
\begin{subfigure}[b]{0.95\linewidth}
   \centering
   \includegraphics[width=1\linewidth]{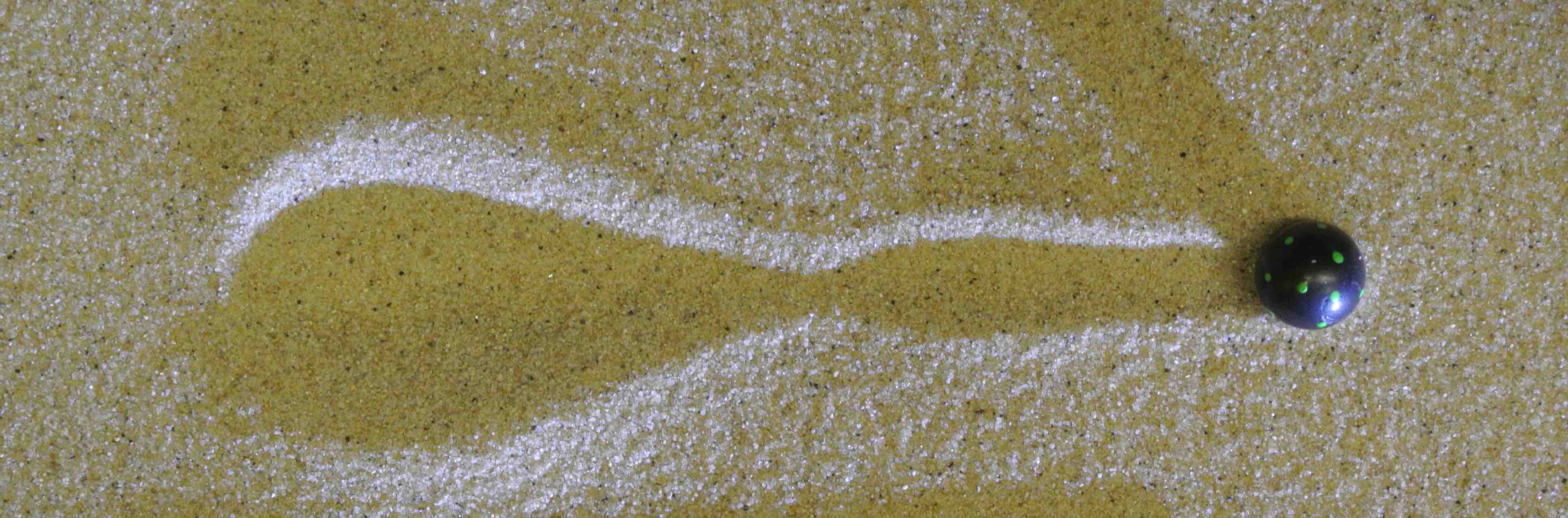}
   \caption{Roll-out}
   \label{fig:crater_rollout}
\end{subfigure}
\begin{subfigure}[b]{0.95\linewidth}
   \centering
   \includegraphics[width=1\linewidth]{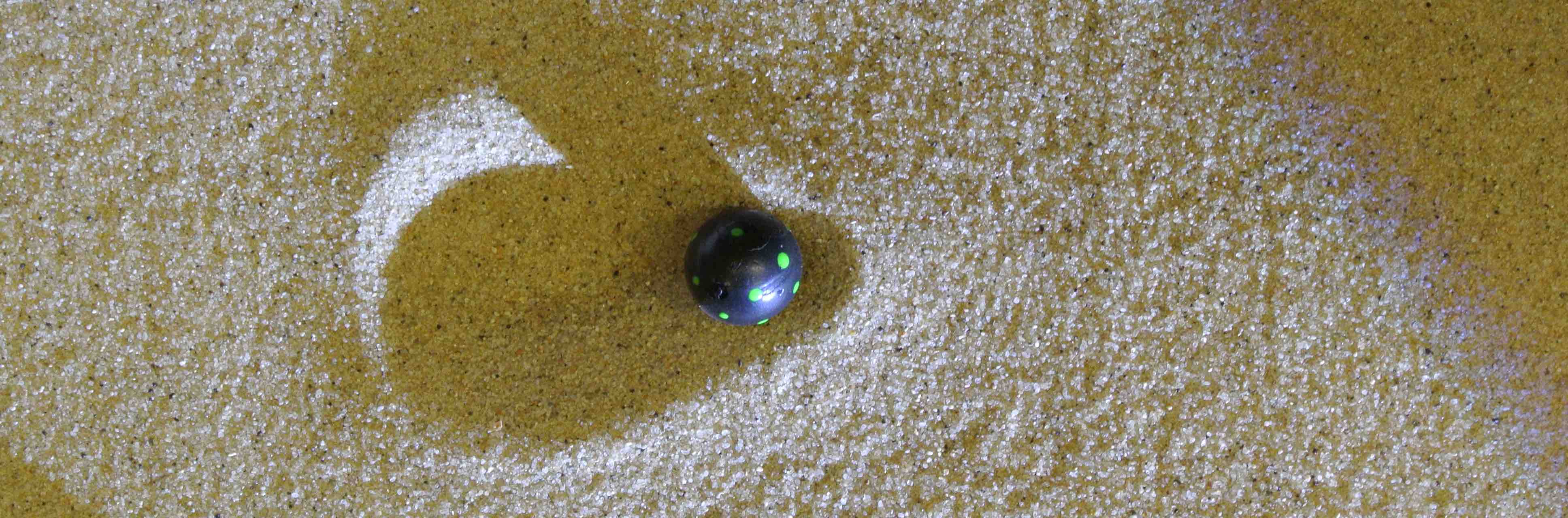}
   \caption{Stop}
   \label{fig:crater_stop}
\end{subfigure}
\caption{We show three different types of impact craters from our experiments. 
These are photographs taken from above the sandbox after three impact experiments. The top photo shows a ricochet, the middle one a roll-out and the bottom one a stop event. Ricochet events have a clear gap between the primary and secondary craters when the marble was above the sand. A roll-out event is one where the marble rolled out of the primary crater, but never lost contact with the sand. The stop event occurs when the marble remains within its primary crater. The marble shown has a diameter of 16.15 mm.
Ricochets and roll-outs tend to occur at higher velocity and lower (or grazing) impact angles.}
\label{fig:craters}
\end{figure}

% ejecta still images
% \begin{figure}
% \centering
% \begin{subfigure}[b]{0.95\linewidth}
%   \centering
%   \includegraphics[width=1\linewidth]{pend030_ejecta.png}
%   \caption{Ricochet}
%   \label{fig:ejecta_ricochet} 
% \end{subfigure}
% \begin{subfigure}[b]{0.95\linewidth}
%   \centering
%   \includegraphics[width=1\linewidth]{pend033_ejecta.png}
%   \caption{Roll-out}
%   \label{fig:ejecta_rollout}
% \end{subfigure}
% \begin{subfigure}[b]{0.95\linewidth}
%   \centering
%   \includegraphics[width=1\linewidth]{pend034_ejecta.png}
%   \caption{Stop}
%   \label{fig:ejecta_stop}
% \end{subfigure}
% \caption{We show ejecta curtains during three different impact experiments.  These three images are from the same experiments shown in Figure \ref{fig:craters}.  These images are individual frames taken from the three high speed videos.}
% \label{fig:ejecta}
% \end{figure}

\subsection{Delineating ricochets from roll-outs and stops}
\label{sec:delin}

Photographs of impact craters from three different laboratory experiments are shown in Figure \ref{fig:craters}.
These photographs were taken from above the sand tray looking downward after the impact of a projectile with velocity of a few m/s and at oblique angles.
We used the resulting impact crater morphologies to classify our impacts as a {\it ricochet}, {\it roll-out}, or {\it stop}.
A crater was classified as {\it ricochet} if there was a clear
gap between a primary crater and a secondary one, as shown in Figure \ref{fig:crater_ricochet}.
An impact was classified as a {\it roll-out} if the marble
rolled out of its crater, as shown in Figure \ref{fig:crater_rollout}.
An impact was classified as a {\it stop} if the marble rested inside
its impact crater after impact, as shown in Figure \ref{fig:crater_stop}.

\begin{figure}
%\centering
\includegraphics[width=3.3in,trim=0 0 0 0]{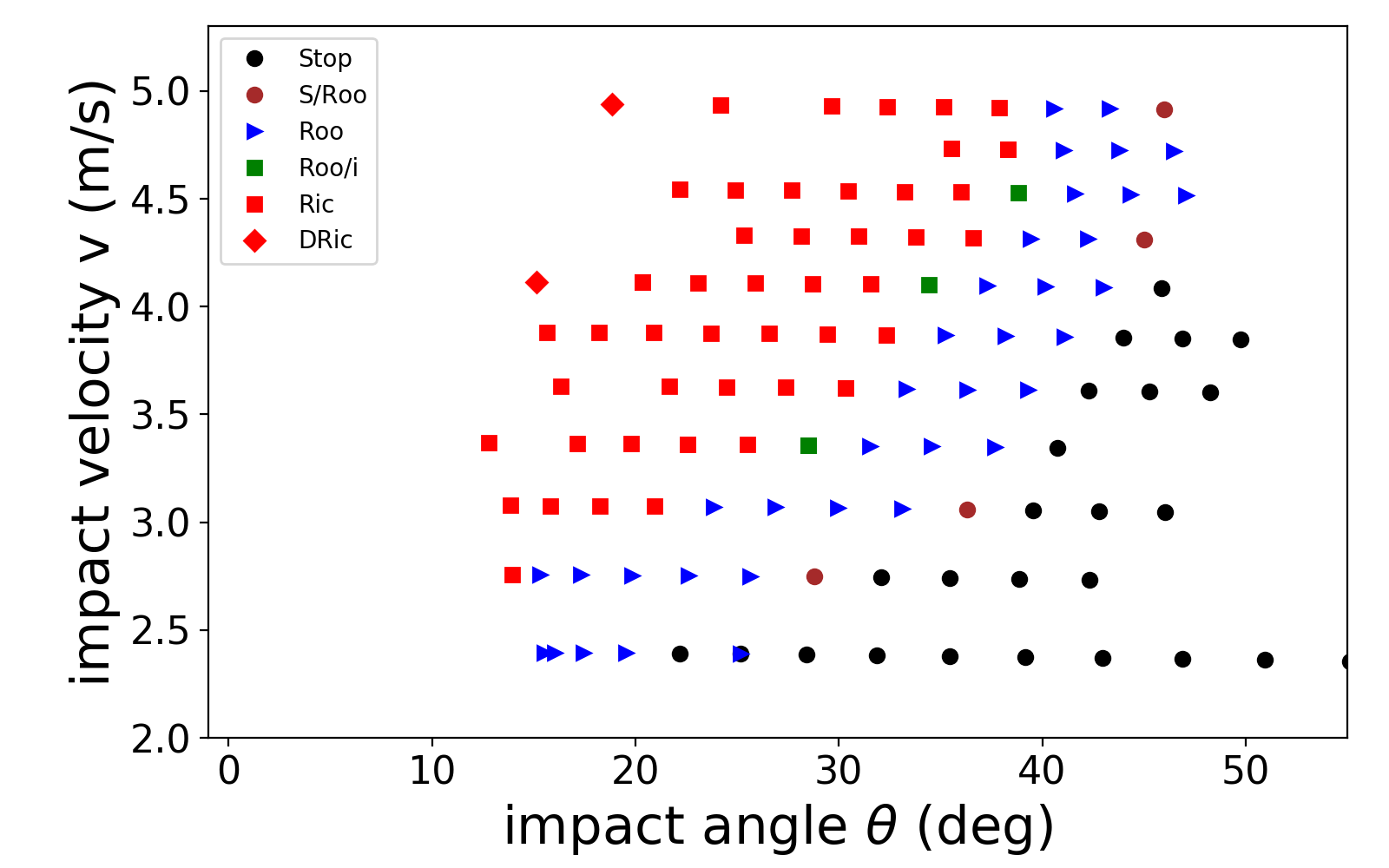}
\caption{Classified impacts as a function of impact angle
and velocity.  Impact angle is measured from horizontal so low $\theta_{impact}$ is a grazing impact.  
Black circles, denoted `Stop' in the legend, are impacts where the marble stayed within its impact crater.
Blue triangles, denoted `Roo' in the legend are roll-outs.  
Red squares (Ric) are ricochets. Green squares (Roo/i) are on the dividing line of ricochet and roll-out.  Brown circles (S/Roo) are on the dividing line of stop and roll-out. Some of the ricochets bounced twice and are labelled with a red diamond.
Each point represents a single impact trial using the pendulum launcher.  \label{fig:ric}
}
\end{figure}

Using our pendulum setup described above we sampled the 
impact parameter space using about 120 unique combinations of the impact velocity and angle.
The outcomes of the impact experiments are categorized as {\it ricochet}, {\it roll-out}, or {\it stop} and are plotted with different point types and colors in Figure \ref{fig:ric} as a function
of computed $v_{impact}$ and $\theta_{impact}$.
Each combination of impact velocity and angle was run multiple times to check for consistency in results. 
We set the impact velocity by adjusting the height of the pendulum and impact angle by the stop-bar position.
For each impact, we recorded the initial pendulum drop height $h$ and the angle of the stop-bar $\theta_{sb}$. 
These two measurements were used to estimate the velocity of impact $v_{impact}$ and the angle of impact $\theta_{impact}$, measured from horizontal so that a grazing impact has a low impact angle.

We describe how impact velocity and impact angle are computed
from the initial pendulum drop height $h$ and stop-bar angle $\theta_{sb}$.
The initial drop height from the mechanism base, $h$,  and distance of pendulum top to mechanism base
when vertical, $d_{base}$, as shown in Figure \ref{fig:pend}, are used to calculate the initial angle of the pendulum $\alpha$, 
(measured from vertical) with
$\cos \alpha = 1 - \frac{h-d_{base}}{L}$ and using the pendulum length $L$.
The pendulum's angular velocity at the moment the pendulum stops at the stop-bar and the projectile is ejected from its holder is
\begin{equation}
 \dot \theta = 
   \sqrt{2 \frac{M_{pend} }{I_{pend}} g R_{cm} 
     (\cos \theta_{sb} - \cos \alpha) }.
\end{equation}
The radius of the pendulum's center of mass is $R_{cm}$ (from its pivot) and the angle $\theta_{sb}$ is the angle set by the stop-bar from vertical.
The speed of the marble when ejected from its holder is
\begin{equation} 
    v_{eject} = L_m \dot \theta. 
\end{equation}
Here $L_m$ is the radial distance from the pivot to the center of the marble when it is in the launcher.
%is the radial distance between marble and the pendulum pivot when the marble is in its holder.
The marble's horizontal and vertical velocity components are
$v_{x,eject} = v_{eject} \cos \theta_{sb} $ and 
$v_{z,eject} = v_{eject} \sin \theta_{sb} $.

We correct for the distance of projectile free fall before hitting the substrate surface, even though this correction is usually small.
After it is ejected, the marble freely falls a distance of
\begin{equation} 
   dz = L + d_{base} - d_s - L_m \cos\theta_{sb} 
\end{equation}
to hit the sand, where $d_s$ is the height of sand surface above the pendulum mechanism base.
The estimated velocity of projectile impact with the sand substrate is
\begin{equation} 
    v_{impact} = \sqrt{v_{z,eject}^2 +  2 g dz +  v_{x,eject}^2}
    \label{eqn:v_i}
\end{equation}
and the angle of impact (measured from horizontal) is
\begin{equation}
\theta_{impact}={\rm arctan}\left( \frac{\sqrt{v_{z,eject}^2 +  2 g dz}}{v_{x,eject}}\right). \label{eqn:theta_i}
\end{equation}

The velocity and angle of impact along with classifications based on impact crater morphology were used to make Figure \ref{fig:ric}.
Figure \ref{fig:ric} shows that crater morphology and impact behavior depends on both impact angle and
velocity.  At higher velocities and lower impact angles ricochets are more likely.  Below a velocity of about 2 m/s
grazing impacts had projectiles that rolled out of their crater
rather than bounced off the sand. 
The dividing line between ricochet and roll-out and that between ricochet and stop was sometimes sensitive to vibrations and wobble in the apparatus.   
We noticed that the location of the ricochet/roll-out line shifted when
we inserted shims into the pivot holder to keep it from vibrating during impact.  With vibrations reduced,
events were repeatable, with series of three or four trials at the same initial pendulum height and stop-bar position
giving the same impact crater morphology and event classification. 
We tentatively assign
a $\pm 5$\textdegree{} error to each point in Figure \ref{fig:ric} that is due to variations in stiffness in the mechanical launch mechanism that we have tried to minimize. This value was the largest discrepancy between the estimated and measured impact angles.
% Below (in section \ref{sec:error}) we discuss variation between predicted and estimated impact angles and velocities measured from tracking the high speed videos.
% There is some discrepancy between estimated and measured impact angles, but 
Estimated and measured impact velocities agree.
The discrepancy in impact angle is discussed below.

Figure \ref{fig:ric} shows that 
the dividing line between different outcomes is quite sensitive to the angle of impact. 
The dividing line trend is opposite to that found by \citet{soliman76} as we see the critical grazing impact angle dividing ricochets from roll-outs increases as a function of impact velocity, rather than decreases.
There are some differences between our experiments and theirs that might explain this difference in behavior.  
Our projectile density is similar to the substrate density (glass marbles into sand), whereas \citet{soliman76} used denser projectiles (steel, aluminum and lead balls into sand).   Their projectiles have higher 
velocity  (theirs were up to 180 m/s and ours are below 5 m/s).    
Both our and their projectiles are spherical.  
Both sets of experiments can be
considered at high Froude number where impact velocity gives 
$Fr  > 1$. The Froude number
of our impacts are 10--20, whereas those by \citet{soliman76} are 300--600.
% 350-650 ft/s and 1 in shot diameter which is 1.27 cm

Our projectiles remain in a shallow penetration regime, where the maximum penetration depth rarely exceeds
the projectile diameter.  We designed our experiment to minimize initial projectile spin. Unfortunately projectile spin is not discussed by these early works (though see the discussion on bouncing bombs by \citealt{johnson98}).
Models that predict granular flow above a particular stress level \citep{bagnold54}
might account for the higher lift we infer in our lower velocity experiments that give us ricochets at high impact angles, as the medium could be effectively stiffer at lower impact velocity. Alternatively lower velocity and lower projectile density, giving shallower levels of penetration, may be increasing the likelihood of ricochets in our experiments.   Since collisions on asteroids are likely to have similar projectile  and substrate densities, our experiments suggest that ricochets could be common in the low velocity regime.

\subsection{The critical angle for ricochet}
\label{sec:crit}

A ricochet takes place when the lift force is at least large enough to overcome the gravitational force
and this must happen before drag forces reduce
the horizontal velocity component velocity to zero \citep{johnson75,soliman76,bai81}.
For spherical projectiles,  a lift force dependent on the square of the depth times the square of the velocity and a constant 
downward gravitational acceleration were adopted by \citet{soliman76} 
to estimate a critical impact angle for ricochet
\begin{equation}
 \theta_{cr}^2 \sim \frac{1}{10} \frac{\rho_s}{\rho_p} - \frac{4 R_p g}{v_{impact}^2} .
 \label{eqn:soliman}
 \end{equation}
At a particular velocity,  ricochets occur at impact angles below the critical one.
The  term on the right is proportional to the inverse of the square of the Froude number and was originally calculated for ricochets on water \citep{johnson75,birkhoff44}.
The term on the left is only dependent on the substrate and projectile density ratio.
At high velocities the critical angle only depends on the density ratio.
In the limit of high velocity, equation \ref{eqn:soliman} predicts a critical angle of $15^\circ$ for our substrate to projectile density ratio of $\rho_s/\rho_p \sim 0.66$.
At the higher velocities in Figure \ref{fig:ric}, we saw ricochets at impact angles up to $40^\circ$, so this model does not apply very well in the regime of our experiments.

Equation \ref{eqn:soliman} predicts  that the critical angle is larger
at higher velocity.  This behavior is seen on water, but ricochets on sand can deviate from this behavior and can scale in the opposite way with critical angle decreasing at higher velocity (see Figure 8 by \citealt{soliman76}).
\citet{bai81} modified equation \ref{eqn:soliman} with the addition of a constant pressure term
dependent on parameter $K'$
adding to the lift,
\begin{equation}
 \theta_{cr}^2 \sim \frac{1}{10} \frac{\rho_s}{\rho_p} - \frac{4 R_p g}{v_{impact}^2}  + \frac{K'}{v_{impact}^2}.
 \label{eqn:bai}
 \end{equation}
This additional term allowed them to account for a 
decreasing critical angle with increasing impact velocity, which was seen in their experiments.

\begin{figure}
%\centering
\includegraphics[width=3.3in,trim=0 0 0 0]{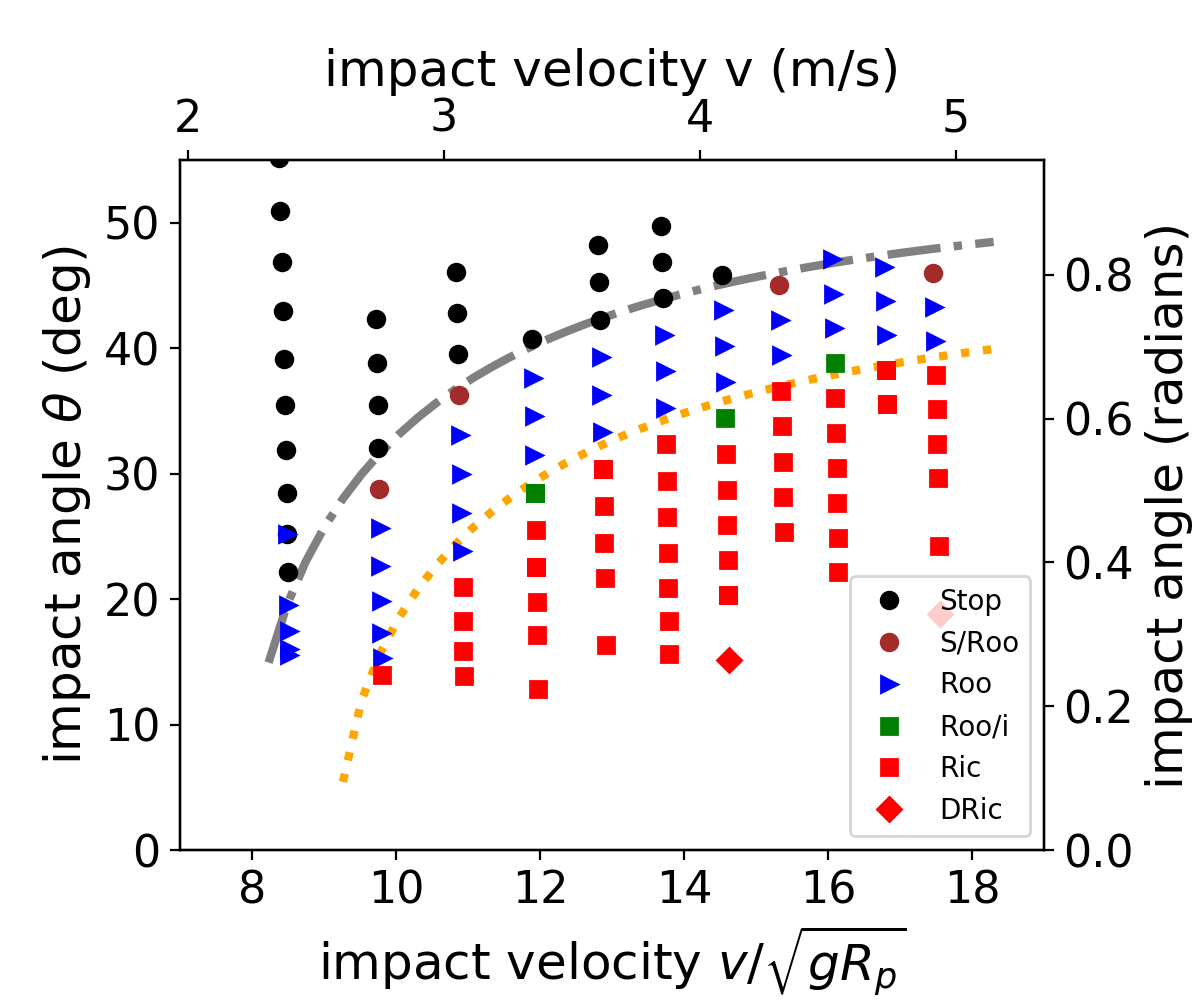}
\caption{The dots are the same events from experiments shown Figure \ref{fig:ric}.  Here the  lower $x$ axis
is the impact Froude number or the
 impact velocity in units of $\sqrt{g R_p}$ where $R_p$ is the marble radius. 
 The top axis is velocity in $m/s$.
The left $y$ axis is the grazing impact angle in degrees and that on the right in radians. The orange dotted line shows
equation \ref{eqn:orange} which is in the form of equation \ref{eqn:soliman} (based on that by \citealt{soliman76})  
but with larger coefficients.  The gray dot-dashed line is equation \ref{eqn:gray} and approximately separates
the roll-outs from the stop events.
 \label{fig:ric_flip2}
}
\end{figure}

As we see an increase in impact angle with increasing velocity, the simpler model by \citet{soliman76} might give a line that matches the division seen in our experiments.
We found a similar line that does delineate the impact outcomes and it is shown in  Figure \ref{fig:ric_flip2}
as a dotted orange line.  Figure \ref{fig:ric_flip2} shows the same experiments as Figure \ref{fig:ric} except
we rotated the axes so that $\theta_{cr}(v_{impact})$ is a function of the 
%impact velocity.
%The lower $x$ axis is the 
Froude number or $\bar v_{impact} = v_{impact}/\sqrt{g R_p}$ with $R_p$ the marble radius.
The orange dotted line that separates the ricochets from the roll-outs is 
\begin{equation}
 \theta_{cr,ric}^2 = 0.65 - \frac{55}{\bar v^2} 
 \label{eqn:orange}
 \end{equation}
 with $\theta_{cr,ric}$ in radians.  A similar line, 
 separating roll-outs from stops is shown as a grey dot-dashed line on Figure \ref{fig:ric_flip2}.
 \begin{equation}
 \theta_{cr,roo}^2 = 0.88 - \frac{55}{\bar v^2} .
 \label{eqn:gray}
 \end{equation}
 
The second term in equation \ref{eqn:soliman} is 4 times the square of the Froude number
but our orange dotted line requires a number 13 times larger than this. 
%In the model by \citet{soliman76}, this term arises from a ratio of gravitational and lift forces.  The large size might be consistent with upward velocity independent  hydrostatic-like forces seen in normal impact experiments that are 5-10 times larger than expected for hydrostatic pressure and Coulomb friction  \citep{katsuragi07,katsuragi13}.  
The constant term  in Equation \ref{eqn:soliman} was predicted 
%in equation \ref{eqn:soliman} 
to be 1/10th the density ratio.  
Our substrate to marble density ratio is about 0.64
so the size of the constant term in Equation \ref{eqn:orange} is about 10 times higher than expected.  
The orange dotted line on Figure \ref{fig:ric_flip2} (from equation \ref{eqn:orange}) is
not consistent with  the ricochet model by \citet{soliman76}.

The orange dotted line on Figure \ref{fig:ric_flip2} (equation \ref{eqn:orange}) represents our  first attempt to 
model the  line dividing ricochets from other types of events.    
% After discussing measurements made from tracking the projectile motion
% from the high speed videos, we make another attempt to create an empirical model for ricochet/roll-out and roll-out/stop dividing lines that are present in Figure \ref{fig:ric} and \ref{fig:ric_flip2}.
This expression will be revisited and improved later on in this paper only after the experimental measurements are discussed.

\section{Trajectories}

The marbles were painted with a black undercoat.  On top of the undercoat they were painted with 18 dots of fluorescent paint to aid in tracking the projectile's spin.  We lit the experiment with bright blue LEDs, causing the paint dots to fluoresce green.  
The LEDs are CREE XLamp XT-E Royal Blue that peak at 450\ nm.\footnote{\url{https://www.cree.com/led-components/media/documents/XLampXTE.pdf}}
We were careful to use a fluorescent paint that is detected as bright when viewed with our high speed video camera. We used 4 wide angle blue LEDs to light the sandbox, primarily from the right side, as shown in Figures \ref{fig:pend}  \ref{fig:ric_lighting}, and \ref{fig:setup}.  The 4 blue light sources made this lighting fairly diffuse.  %We found that our camera is insensitive to common red fluorescent paints and that different shades of fluorescent blue paints were indistinguishable.

We also lit the impact region with a single bright white light from the top left side (see the photograph in Figure \ref{fig:setup}). With projectiles moving left to right, this gave a single white reflection on the marble that could be seen from the front of the experiment during most of the impact.   The ejecta curtain tended to obscure the right side of the marble during the impact. We used the white reflection to track the marble's center of mass motion.

We filmed the impacts with a Krontech Chronos 1.4 high speed camera at 3000 frames per second.  The marble diameter (16.15 mm) was used to find the pixel scale in the video image frames.
The high speed videos were taken from a 45$^\circ$ angle from vertical (see Figure \ref{fig:setup}), allowing us to track both horizontal and vertical projectile motions.
The impact craters were used to verify that marble trajectories remained in the pendulum's plane. 
Videos used to track the projectiles can be found in the supplemental materials.

%\FloatBarrier

\subsection{Data reduction}
\label{sec:data}

As our high speed camera takes color images, using weighted sums of red, green and blue color channels we could emphasize the white light reflection or remove it and focus on the fluorescent green markers.   The white light reflections were used to track the marble center of mass. The fluorescent green markers were used to measure the marble spin.

We use the soft-matter particle tracking software package \texttt{trackpy} \citep{trackpy} to identify and track the fluorescent dots and reflections on the projectile seen in individual video frames.
\texttt{Trackpy} is a software package for finding blob-like features in video, tracking them through time, linking and analyzing the trajectories. It implements and extends in Python the widely-used Crocker-Grier algorithm for finding single-particle trajectories \citep{crocker96}. 

\begin{figure*}
\centering
\begin{subfigure}{\linewidth}
   \centering
   \includegraphics[width=\linewidth,trim = {0 50 0 80}, clip]{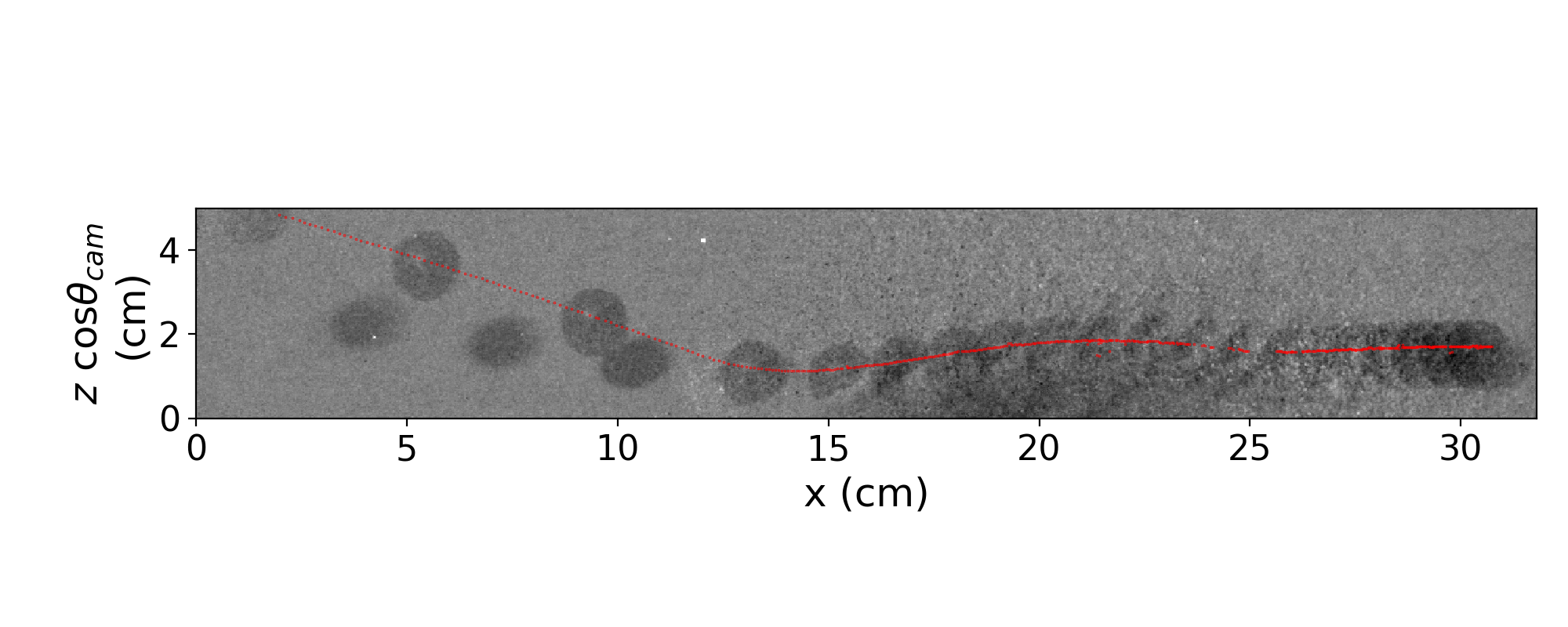}
   %\caption{Ricochet event}
   \label{fig:rico_seq} 
\end{subfigure}
\begin{subfigure}{\linewidth}
   \centering
   \includegraphics[width=\linewidth,trim = {0 30 0 80}, clip]{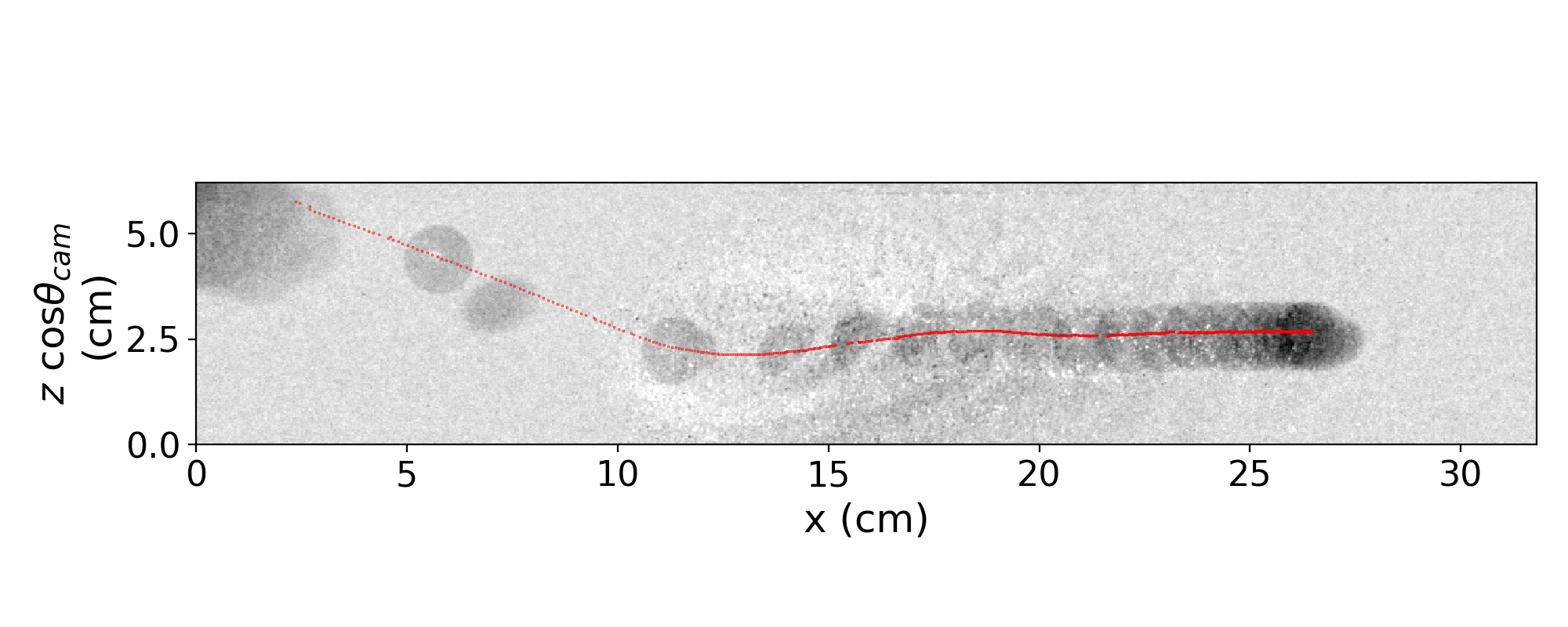}
   %\caption{Roll-out event}
   \label{fig:rollout_seq}
\end{subfigure}
\begin{subfigure}{\linewidth}
   \centering
   \includegraphics[width=\linewidth,trim = {0 40 0 80}, clip]{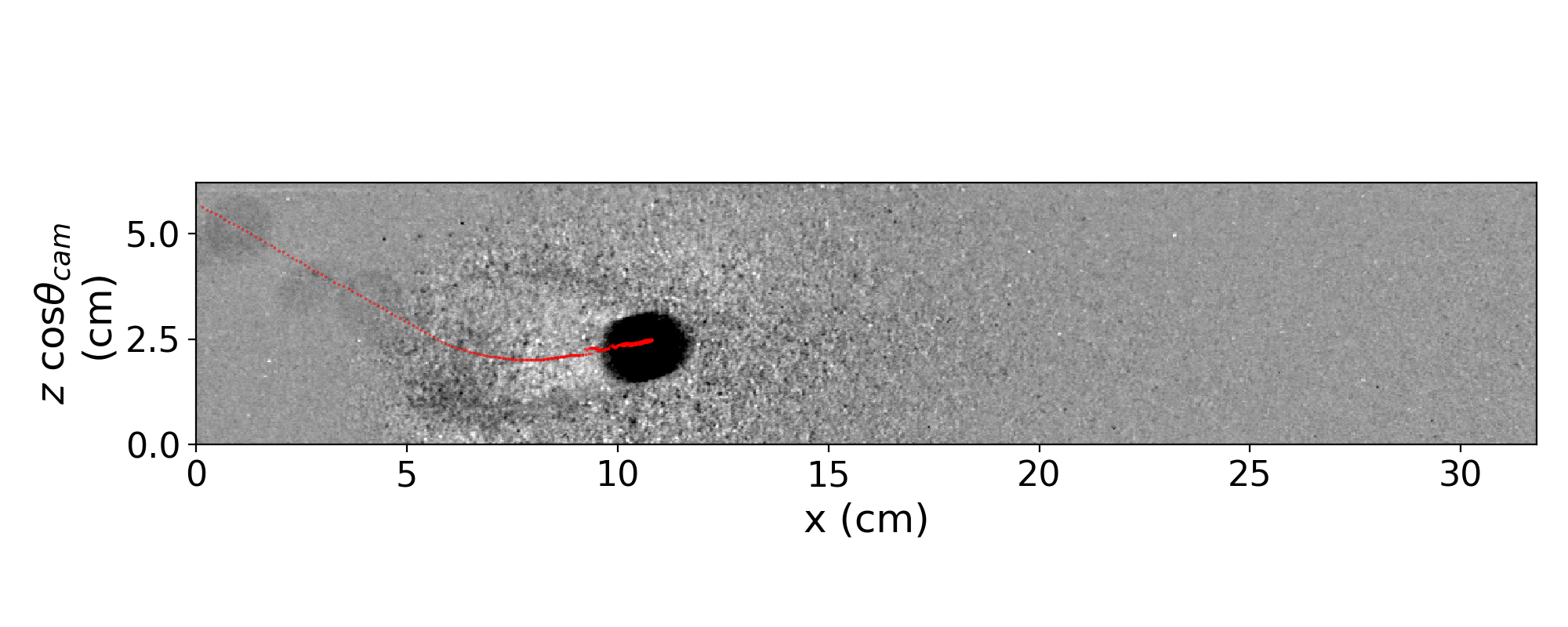}
   %\caption{Stop event}
   \label{fig:stop_seq}
\end{subfigure}
\caption{Tracking a white light reflection on the marble to measure its trajectory.   Trajectories are shown
for three different high speed videos. The horizontal axis is the marble's x-position. 
The vertical axis is the height of the marble with the camera pointed at a 45\textdegree{} angle with respect to the pendulum plane (xz-plane).
We show in grayscale a sum of high speed video images that are separated by 0.01s.
The red line is the center of mass track that goes through a white light reflection seen on the top left side of the marble.  The marble started on the upper left and came to rest outside the field of view on the right.
The marble is partially obscured by the ejecta curtain during part of its trajectory.  In the top panel the marble ricocheted off the sand in the middle of the figure and bounced upward, then hit the sand again and rolled across it to the right. The middle panel shows the marble rolling out of its initial crater then coming to rest. The bottom panel shows the trajectory of the marble stopping within its impact crater.  At the end of the video, the marble rolled backwards back  down into its crater.}
\label{fig:CM_seq}
\end{figure*}

We first tracked the position of the white light reflection on the marble in each video frame.  Prior to tracking we used a series of images to construct a median image which was subtracted from each video frame. 
We adjusted the radius and integrated peak brightness so that the reflection was identified by the tracking software and also so that the number of sand particles tracked is reduced.
We eliminated spurious tracks by hand, leaving only the tracked white light reflection.  
The white light reflection track measured from three high speed videos is shown on top of a sum of images in Figure \ref{fig:CM_seq}. The horizontal axis is the marble's position in x and the vertical axis is the projected z direction as the camera was positioned 45\textdegree{} above the sand tray's surface plane.

\begin{figure*}
%\captionsetup[subfigure]{justification=centering}
\centering
\begin{subfigure}[t]{0.3\linewidth}
   %\centering
   \includegraphics[width=\columnwidth,trim={10 60 0 0},clip]{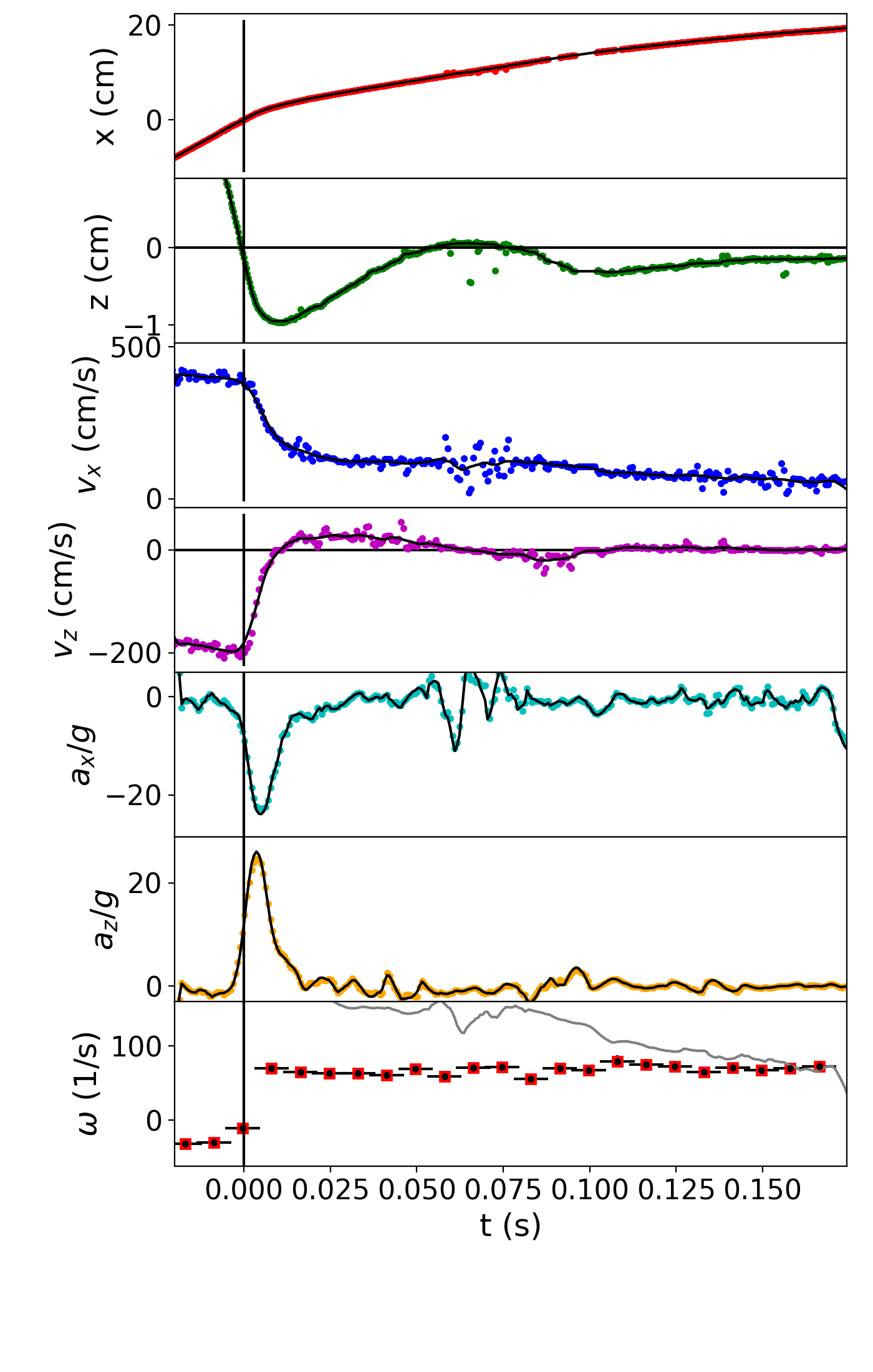}
   \caption{Ricochet}
   \label{fig:rico_traj} 
\end{subfigure}
\begin{subfigure}[t]{0.3\linewidth}
   %\centering
   \includegraphics[width=\columnwidth,trim={5 60 5 0},clip]{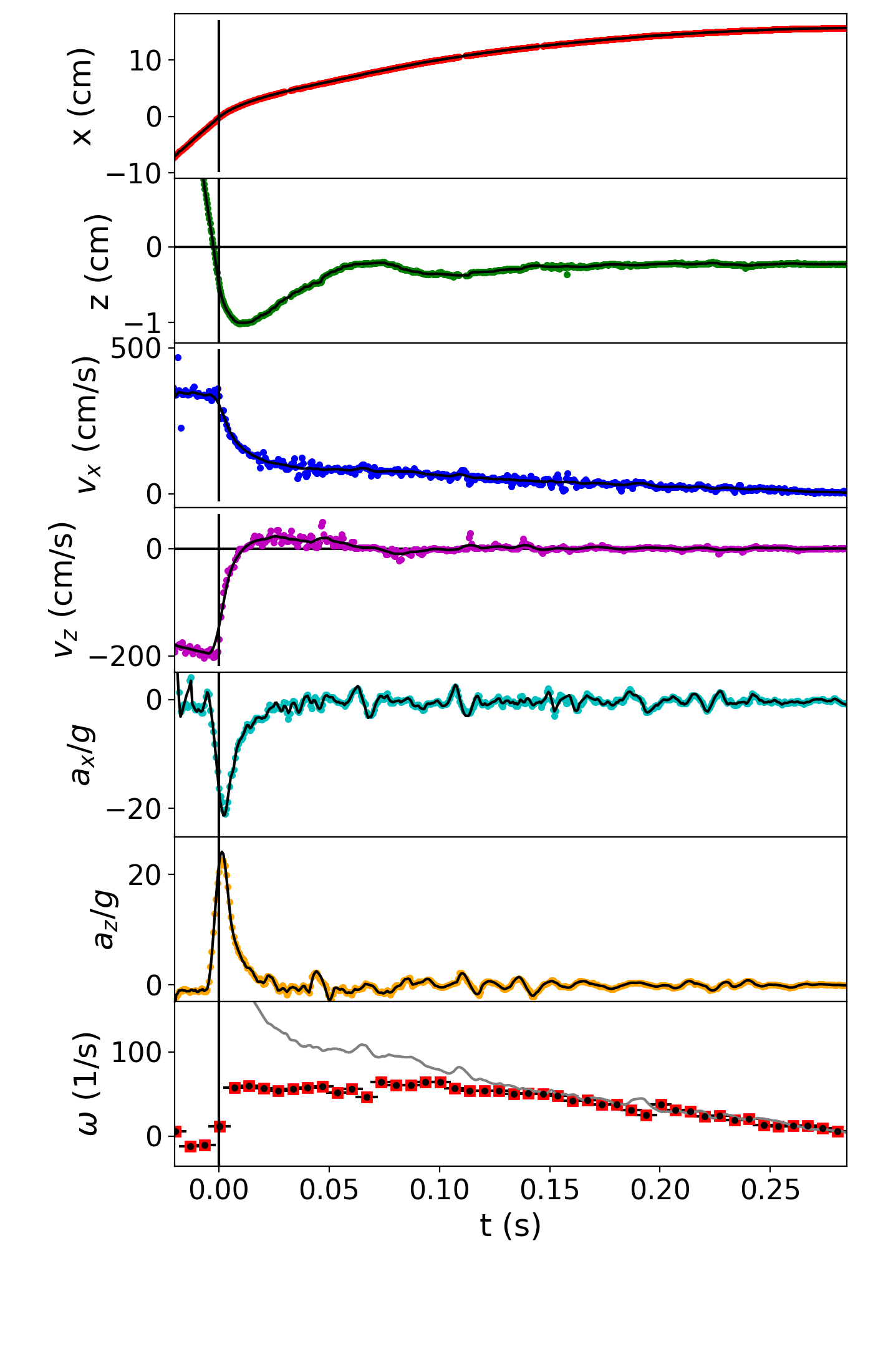}
   \caption{Roll-out}
   \label{fig:rollout_traj}
\end{subfigure}
\begin{subfigure}[t]{0.3\linewidth}
   %\centering
   \includegraphics[width=\columnwidth,trim={0 60 10 0},clip]{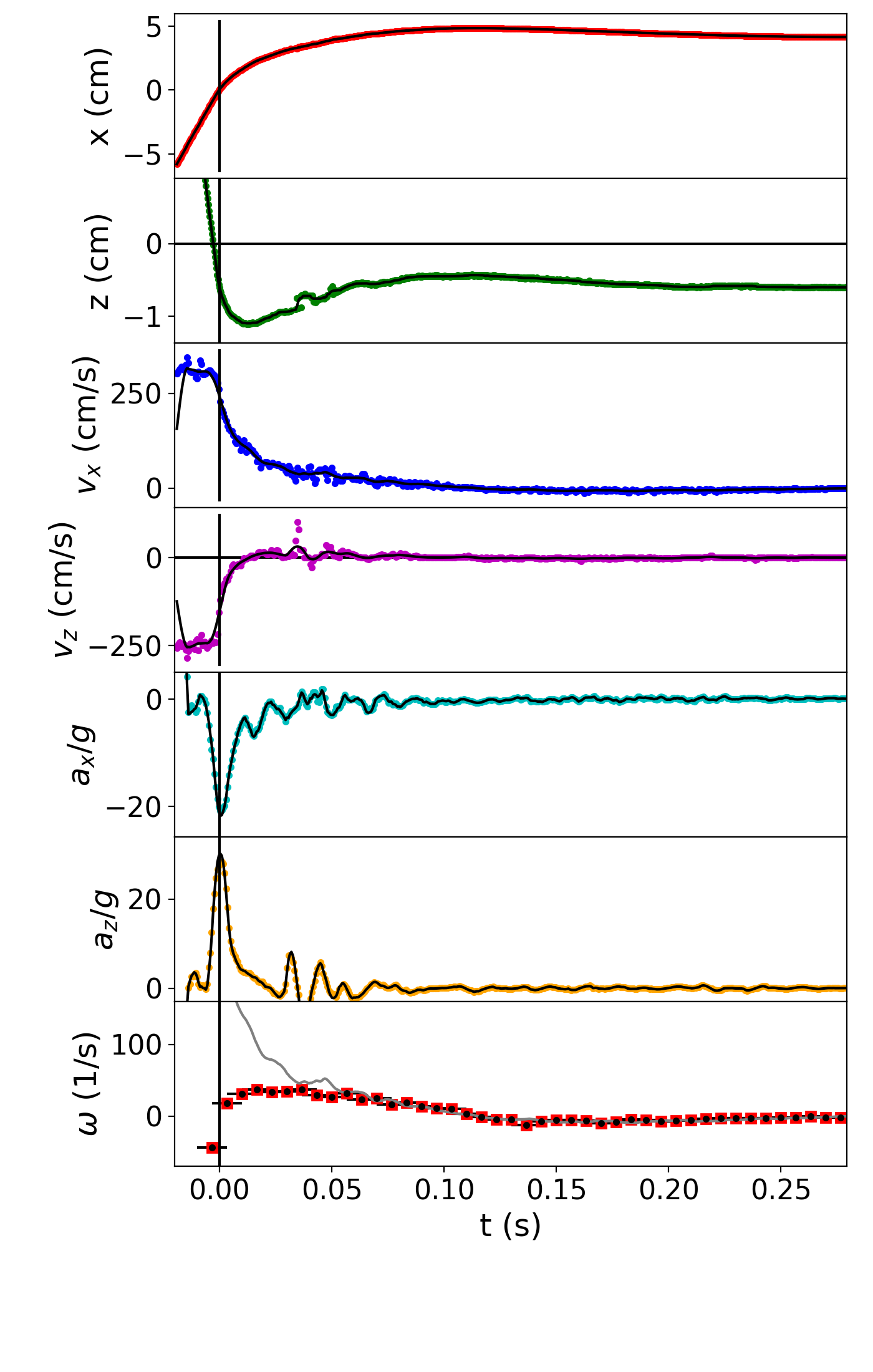}
   \caption{Stop}
   \label{fig:stop_traj}
\end{subfigure}
% \begin{figure*}
% %\centering
% \begin{tabular}{ccc}
% \includegraphics[width=2.2in,trim =  0 0 0 0, clip]{pend030_traj2.png}
% \caption{Roll-out event}
% &
% \includegraphics[width=2.2in,trim = 0 0 0 0, clip]{pend033_traj2.png}
% &
% \includegraphics[width=2.2in,trim = 0 0 0 0, clip]{pend034_traj2.png}
% \end{tabular}
% \\
% a & b & c
\caption{Marble center of mass trajectories for three impact experiments. 
%From the left to right, the videos are pend030, pend033, and pend034. 
The leftmost figure is a  ricochet,
the middle one is a roll-out and the right one is a stop.
We show horizontal and vertical positions, horizontal and vertical velocities and accelerations as a function of time.  
A vertical position of $z>0$ has the bottom of the marble above
the sand.  The $x$ position is measured from the point of impact and with $x$ increasing along the direction of motion.
Estimated time of impact is shown with the black vertical line.
Colored dots show coarser measurements of the trajectory positions, 
velocities and accelerations.  
Black lines show median filtered and smoothed versions.
For the spin in the bottom panel, the horizontal error bars show the intervals used to measure the spin.
The gray lines in the lower panel are $v_x/R_p$.  When the grey lines  lie are near the spin measurements, the marble is rolling without slipping.
The experimental settings of these videos is given in Table \ref{tab:exp_list}.
\label{fig:pend_traj}
}
\end{figure*}

The positions of the white light reflection plus a constant offset gives us an estimate for the marble center of mass position as a function of time.  We adopt a coordinate system with $x$ increasing
in the horizontal  direction along the direction of the projectile motion and $z$ increasing in the vertical direction.  Projectile trajectories remained nearly in the $xz$ plane (pendulum plane).
The $x$ and $z$ marble position vectors were interpolated from the arrays of tracked positions so as to be evenly sampled in time.
The vertical positions were corrected to take into account the camera viewing angle from horizontal.  
Because the camera frame was oriented parallel to the horizontal direction of the projectile motion, we did not need to correct the $x$ direction for camera viewing angle.
The time vector is computed from frame numbers by dividing by the video frame rate (3000 fps).
We estimated the time and position of impact from the first frame showing an ejecta curtain.
We median filtered the $x$ and $z$ position arrays using a width of 11 samples which is 3.6 ms at a sampling rate of 3000 Hz.  To compute velocities and accelerations, we smoothed the arrays using a Savinsky-Golay filter with widths of 15 and 17 samples respectively.
We checked that the white light reflection used to track the center of mass of the marble did not change position on the surface during its motion. 
%See Figure \ref{fig:reflection} in the appendix.
Trajectories of the marble center of mass as a function of time are shown in Figure \ref{fig:pend_traj}. The origin of these plots correspond to the time and location of impact.

\begin{figure*}
\centering
\begin{tabular}{ccc}
\includegraphics[width=2.0in,trim =30 0 20 0, clip]{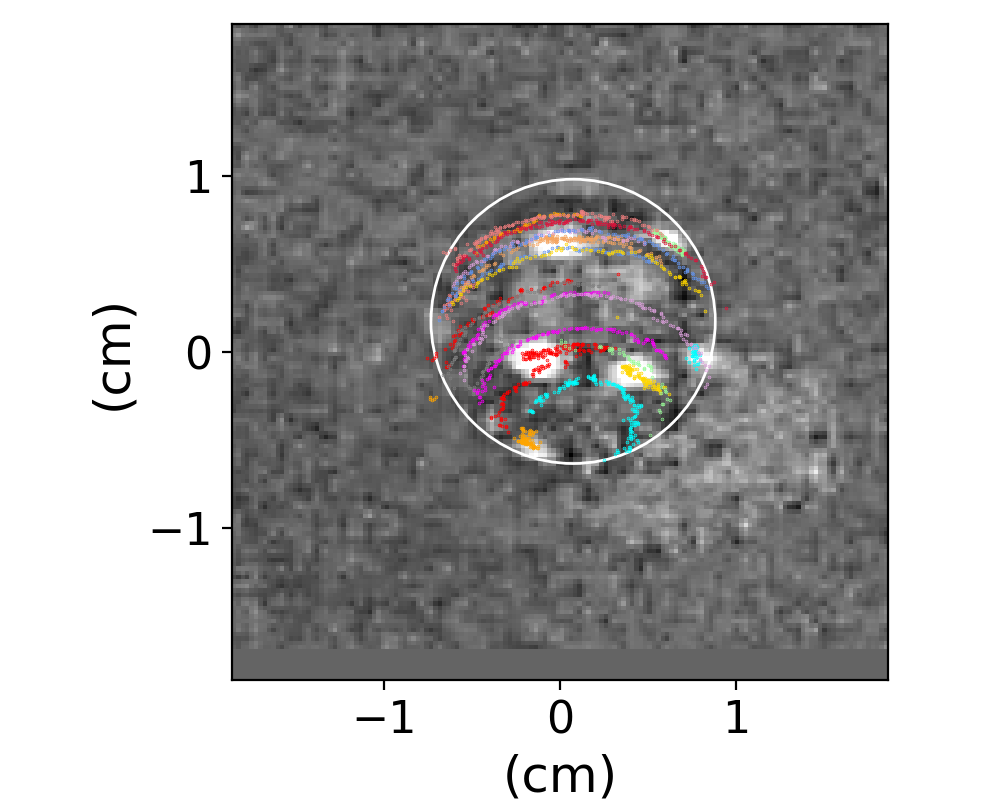} &
\includegraphics[width=2.0in,trim =  30 0 20 0, clip]{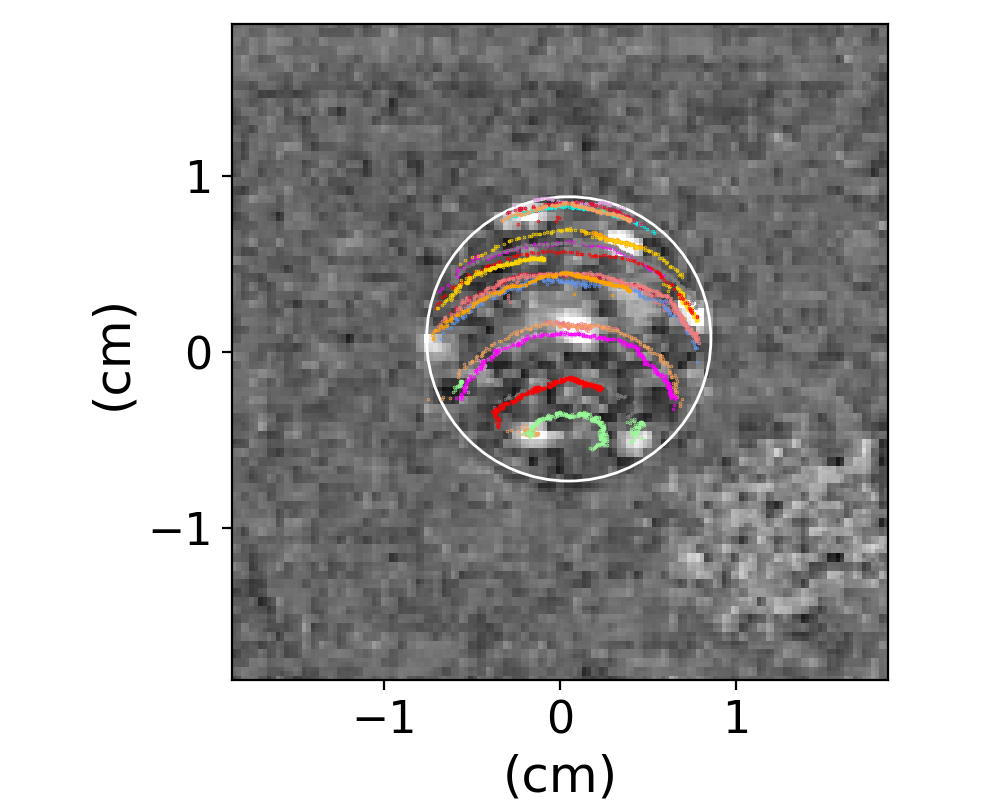} &
 \includegraphics[width=2.0in,trim =  30 0 20 0, clip]{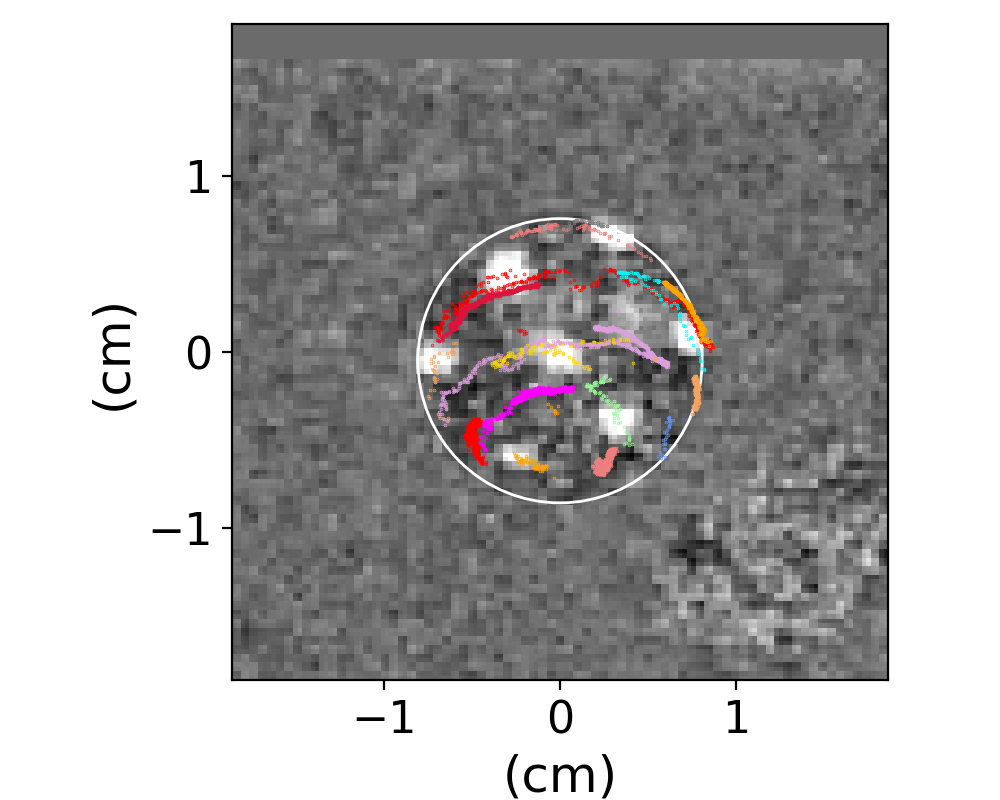}
\end{tabular}
\caption{We show tracks of the fluorescent green markers in the center of mass frame overlayed on top of a single frame from the high speed video that is shown in gray-scale. 
From left to right, we show tracks for the Ricochet, Roll-out, and Stop videos.
In each figure, the tracks were taken from times spanning the video and each track is shown in a different color.  
The tracks are consistent with the $45^\circ$ camera angle
 and a spin vector that did not vary during the impact and
 maintained orientation parallel to the substrate surface.
\label{fig:spintracks}
  }
\end{figure*}

To measure the marble spin we used the video frames' green channel, showing the
fluorescent markers.
We shifted each image using the previously computed marble center of mass positions to put the marble in the center of the image.  We then tracked the fluorescent markers again using the \texttt{trackpy} software package. 
Tracks of the fluorescent markers spanning whole videos are shown in Figure \ref{fig:spintracks} for three videos.  Even though the marble spin varies as a function of time, the tracks in each figure lie on similar arcs.  This implies that the spin orientation did not significantly vary throughout the video.
This does not imply that the marble's angular rotation rate remained fixed.
The orientation of the arcs are consistent with a horizontal spin axis and the camera orientation angle, $\theta_{cam}$, of 45\textdegree{} with respect to vertical.

To measure the spin angular rotation rate we fit tracks during short intervals of time ($\sim10$ ms). Assuming a spherical surface, each track is described by the spin orientation angle, the angular rotation rate and an initial fluorescent dot position on the marble surface.  We constructed a minimization function that is the sum of differences between predicted (via rotation) and observed tracked particle positions.  
We simultaneously fit for the initial dot positions and the angular rotation rate. %Comparisons between tracks and fit tracks for a series of time intervals are shown in Figure \ref{fig:pend030_fit}. This figure shows that our fit angular rotation rates successfully put modeled tracks on top of those ones observed. 

The vertical error bars for the angular rotation rates were determined by varying parameters that went into the angular velocity fitting routine such as center of mass position, camera angle, and projectile radius. The largest source of error was identified from the uncertainty in the radius. An uncertainty of 3 pixels in the radius gave a 10\% error in the spin values. %shown in Figure \ref{fig:pend030_fit}. 
A radius of 33 pixels was used for our projectile spin fitting.
Horizontal error bars for the angular rotation rates show the time interval used to measure the spin.
%{\bf Discuss spin errorbars. xxxx Also in updated figures}

% \subsubsection{Crater measurements}

% Projectile impacts always formed a crater in the sand. 
% After impact, we turned off the blue LEDs and illuminated the crater with a white LED spotlight at grazing angle from the front. 
% We took photos to measure the crater extent and used a ruler to measure its maximum depth with respect to the level of undisturbed sand.
% Examples of such photographs are shown in Figure \ref{fig:craters}. 

% {\bf Which table are these measurements put in? xxxx}

%\subsection{Sources of error}
%\label{sec:error}

Table \ref{tab:exp_list} lists experiments along with initial pendulum arm settings: height h, the stopping bar angle $\theta_{sb}$, and distance of sand surface to the pendulum base $d_s$.
Predicted impact velocities $v_{imp}$ and angles $\theta_{imp}$ were found using equations \ref{eqn:v_i} \& \ref{eqn:theta_i} respectively.
Measured impact velocities and angles were obtained from marble trajectories just before impact.
The predicted and measured impact velocities are consistent.
The measured impact angles were about 5\textdegree{} lower than predicted for the ricochet and roll-out events.
%but are consistent with the stopping bar setting.

The discrepancy between the predicted and measured angles are attributable to errors in the pendulum setup.
Soft rubber was used on the launcher to better hold the marble. This could lead to a nonuniform suction causing the marble to not separate from the launcher once it hits the stopping bar. The pendulum hitting the stopping bar also caused the post to bend slightly. This was minimized by using a thicker post. The pendulum arm can also bounce when hitting the stopping bar. Lead weights were added to the base of the pendulum arm to reduce shaking during marble launch.

% trajectory sequences

% Single CoM sequence figure
% \begin{figure*}
% \centering
% \includegraphics[width=6.5in,trim =  30 40 70 90, clip]{pend030_seq.png}
% \caption{Tracking a single white light reflection on the marble to measure its trajectory. 
% We show in grayscale a sum of high speed video images that are separated by 0.01s.
% The red line is  the center of mass track that goes through a white light reflection seen on the top left side of the marble.  The marble started on the upper left and came to rest outside the field of view on the right.
% The marble is partially obscured by the ejecta curtain during part of its trajectory.  The marble ricocheted off the sand in the middle of the figure and bounced upward, then hit the sand again and rolled across it to the right.
%  \label{fig:pend030_seq}
%   }
% \end{figure*}

% \begin{figure*}
% \centering
% \includegraphics[width=6.0in,trim =  90 200 50 150, clip]{pend030_fit.pdf}
% \caption{We tracked fluorescent markers on the marble. These are used to make measurements of the marble angular rotation rates at different times. 
% Each panel shows markers during different time intervals in the high speed Ricochet video. The central time of each interval from impact in seconds is shown on the top right of each panel. Colored points show tracked fluorescent markers. 
% Each marker is shown with a different point color. Black lines shows the fitted rotation model for each tracked fluorescent dot.
% The arrows show the direction of the spin rotation axis.  This direction was  not allowed to vary during fitting, however the angular rotation rate and marker positions in the marble body frame were fit parameters.  
% \label{fig:pend030_fit}
% }
% \end{figure*}

% \begin{figure*}
% \centering
% \includegraphics[width=6.0in,trim =  90 200 50 150, clip]{pend033_fit.pdf}
% \caption{Similar to Figure \ref{fig:pend030_fit} but for a Roll-out.
%  \label{fig:pend033_fit}
% }
% \end{figure*}

% \begin{figure*}
% \centering
% \includegraphics[width=6.0in,trim = 50 270 50 150, clip]{pend034_fit.pdf}
% \caption{Similar to Figure \ref{fig:pend030_fit} but for a Stop.
% \label{fig:pend034_fit}
% }
% \end{figure*}

\begin{table}
{
%\centering
\caption{Quantities}
\label{tab:pend}
\begin{tabular}{lll}
\hline
Length of pendulum   & $L$ & 94.8 cm \\
Period of small oscillations & $T$ & 1.78 s \\
Moment of inertia divided by mass & $\left(\frac{I}{M}\right)_{pend}$ &  3972  cm$^2$\\
Radius of pendulum's center of mass  & $R_{cm}$ & 50.5 cm \\
Radius of marble holder from pivot & $L_m$ & 84.3 cm \\
Distance of pendulum tip to base  & $d_{base}$ & 1.5 cm \\
Distance of base to sand surface & $d_s$ & 9.9 cm\\
\hline
Diameter of marble & $2 R_p$ & 16.15 mm \\
Mass of marble & $m_p$ &  5.57 g \\
Density of marble & $\rho_p$ & 2.5 g/cm$^3$ \\
Unit of velocity & $\sqrt{g R_p}$ & 28.1 cm/s \\
\hline
Density of sand & $\rho_s$ & 1.6 g/cm$^3$ \\
Sand angle of repose & $\theta_s$ & 32$^\circ$ \\
Coefficient of friction & $\mu$ & 0.51 \\
Camera Angle & $\theta_{cam}$ & 45\textdegree{}\\
\hline
Inside dimensions of Sand Tray && 87.5 x 11.5 x \\ 
&& 6.3 cm \\
\hline
\hline
\end{tabular}

Notes: The value for $d_s$ (as shown in Figure \ref{fig:pend}) reported here is for all experiments shown in Figure   \ref{fig:ric}. The coefficient of friction for the sand is computed from its angle of repose $\mu = {\rm atan}(\theta_s)$.} 
\end{table}

\begin{table}
{
%\centering
\caption{Nomenclature}
\label{tab:nomen}
\begin{tabular}{lll}
\hline
Projectile mass & $m_p$  &\\
Surface gravitational acceleration & $g$ &\\
Projectile radius, if spherical & $R_p$ &\\
Granular substrate mean density & $\rho_s$& \\
Projectile density & $\rho_p$ &\\
Projectile velocity at impact  & $v_{impact}$ &\\
Projectile velocity vector & ${\bf v}$& \\
Projectile cross sectional area & $A$ &\\
Critical impact angle & $\theta_{cr}$ &\\
Froude number  & \multicolumn{2}{l}{ Fr $= \bar v = v/\sqrt{R_pg}$ } \\
Horizontal coordinate & $x$&  \\
Vertical coordinate & $z$ &\\
Normalized vertical coordinate & $\bar z = |z| / R_p$ &\\
Depth below surface level  & \multicolumn{2}{l}{ $|z|$ with $z<0$ } \\
Impact angle & $\theta_{impact}$& \\
%Spherical coordinate angles & $\phi,\psi$ \\
Drag force & $F_d$ &\\
Lift force & $F_L$ &\\
Coefficient of static friction  & $\mu_s$ &\\
Angle of repose & $\theta_r$ &\\
Stopping time & $t_s$ &\\
Maximum penetration depth & $d_{mp}$ &\\
Horizontal velocity component  & $v_{xmp}$& \\
~ at maximum depth & &\\
Time of maximum height & $t_m$ & \\
~ during rebound & &\\
Height reached in rebound & $z(t_m)$& \\
Drag coefficients & $\alpha_x, \beta_x $ &\\
Lift coefficient & $c_L$ &\\
Effective friction coefficient & $\mu_{eff}$ &\\
Vertical, horizontal acceleration & $a_z$, $a_x$ &\\
Angular acceleration &  $\dot{\omega}$ &\\
Angle of stop-bar & $\theta_{sb}$ &\\
Height of tip of pendulum arm & $h$ &\\
Drop angle of pendulum arm & $\alpha$ &\\
\hline
\end{tabular}\\

For a grazing impact $\theta=0$.
The vertical coordinate is positive upward.
The horizontal coordinate is positive with the initial direction of projectile motion.}
\end{table}

\begin{table*}
\centering
{
\caption{Video list}
\label{tab:exp_list}
\begin{tabular}{llllllll}
\hline
Event/Video & $h$  & $\theta_{sb}$  & $d_s$  & $v_{imp,predicted}$  & $v_{imp,measured}$  & $\theta_{imp,predicted}$   & $\theta_{imp,measured}$  \\ 
 & (cm) & (deg) & (cm) &(cm/s) & (cm/s) & (deg)& (deg)\\
\hline
Ricochet & 110 & 25 & 9.9 & 454 & 444 & 30.5 & 25.3 \\
Roll-out & 90 & 30 & 14.5 & 399 & 391 & 35.1 & 29.1 \\
Stop & 90 & 35 & 14.5 & 398 & 396 & 38.8 & 38.5 \\
\hline
\end{tabular}}
\\The values for the pendulum, sand and marble, $L$, $L_m$, $d_{base}$, $(I/m)_{pend}$, $m_p$, $\rho_p$, and $\rho_{s}$ are the same for all events and given in Table \ref{tab:pend}.  The predicted impact angle
and velocity were predicted from the initial pendulum height
and stop-bar position using equations \ref{eqn:v_i} and \ref{eqn:theta_i}.
\end{table*}

% \begin{figure*}
% \centering
% \includegraphics[width=6.5in,trim =  30 40 70 90, clip]{pend033_seq.png}
%  \caption{Similar to Figure \ref{fig:pend030_seq} except for a rollout (pend 33).  The marble stopped within the video.
%  \label{fig:pend033_seq}
%   }
% \end{figure*}

\FloatBarrier

\subsection{Shapes of Trajectories}

%\textbf{Discuss Figure of trajectories}

The trajectories shown in Figure \ref{fig:pend_traj} are measured from three different videos. The panels, from left to right, are for a ricochet, a roll-out, and a stop impact event. The top two panels show the x and z position of the white light reflection on the projectile as functions of time. The origin was chosen to be the location and time of impact. The x position increases for all time for all cratering events. The stop event has a decrease in x for late times that corresponds to the marble unable to escape its own initial impact crater and rolling backwards. 

Inspection of the depths as a function of time (second panels from top in Figure \ref{fig:pend_traj}) shows that the marble  bounced above the surface (where $z>0$) during the ricochet event.   The bottom surface of the marble did  rise above the surface level breaking contact with the sand.  The bottom of the marble  remained below surface level after penetration for the roll-out and stop events. 
%It can be seen that our marble never reaches a depth greater than the marble diameter.

In the stop event (rightmost figure in Figure \ref{fig:pend_traj}) and
after maximum penetration depth, the bottom of the marble rose to about $z=-0.5$ cm which places the center of mass of the marble near the surface level.  This can be contrasted with the roll-out event where the bottom of the marble rose to a height
nearly level with the surface and placing the marble's center of mass well above the surface level.  Likely we can think of a  roll-out event as one with depth reached after maximum penetration high enough to put  the center of mass above the surface
and allowing the marble to roll-out of its crater. 

This is consistent with the crater morphology for these events that is shown in Figure \ref{fig:craters} and provides confirmation that our tracking software is working. 
The maximum depth is closer to the point of impact than the crater center giving azimuthally asymmetric shapes  which has been seen in other oblique impact treatments (e.g., \citealt{soliman76,daneshi77,bai81}). 
%By contrast, a normal impact would produce a azimuthally symmetric crater and ejecta curtain.

Prior studies (e.g., \citealt{katsuragi13}) call the maximum depth reached a penetration depth. The penetration depth is reached when the vertical velocity component changes sign. For our tracked videos we have listed measurements at the time of maximum depth in Table \ref{tab:max_pen}.

The trajectories have similar peak acceleration values, with peak vertical acceleration component $a_z \sim 24$ g and peak horizontal acceleration $|a_x|$ slightly less at $|a_x| \sim 22$ g.
\cite{katsuragi07}, \cite{vandermeer17}, and \cite{goldman08} had accelerations that were the same order of magnitude for similar impact velocities. All were normal impact experiments with steel projectiles. 
\cite{vandermeer17} used sand as a substrate whereas the other two used glass beads.

As shown in the bottom panels of Figure \ref{fig:pend_traj}, the marble's spin increases while the deceleration is high right after impact.  On the bottom panel,   the gray lines show $v_x/R_p$ and the red dots show the spin or angular rotation rate. 
When the two coincide, the marble is rolling without slipping. We see that a rolling without slip condition is not reached until later times when the projectiles are at lower velocities.  The marble's surface was moving  with respect to the sand while the ejecta curtains were launched.  The ricochet event (that in the left figure in Figure \ref{fig:pend_traj}) did not achieve rolling without slip until the marble fell back into the sand.  The roll-out event (the middle figure) rolled without slipping past $t \approx 0.14$ s while the marble continued to roll across the sand.

\begin{table}
\centering
{
\caption{At maximum penetration depth}
\label{tab:max_pen}
\begin{tabular}{llll}
\hline
Event/Video & $d_{mp}$ & $v_{xmp}$ & $t_s$ \\
            & (cm)  & (cm/s) & (s) \\
\hline
Ricochet  & 0.95 & 221 & 0.008 \\
Roll-out & 1.00 & 165 & 0.01 \\
Stop  & 1.10 & 108 & 0.012 \\
\hline 
\end{tabular}}
\\ Quantities measured at the moment of maximum penetration from the trajectories plotted in Figure \ref{fig:pend_traj} for the three experiments presented.  
\end{table}

\begin{figure}
\centering
\includegraphics[width=3.2in]{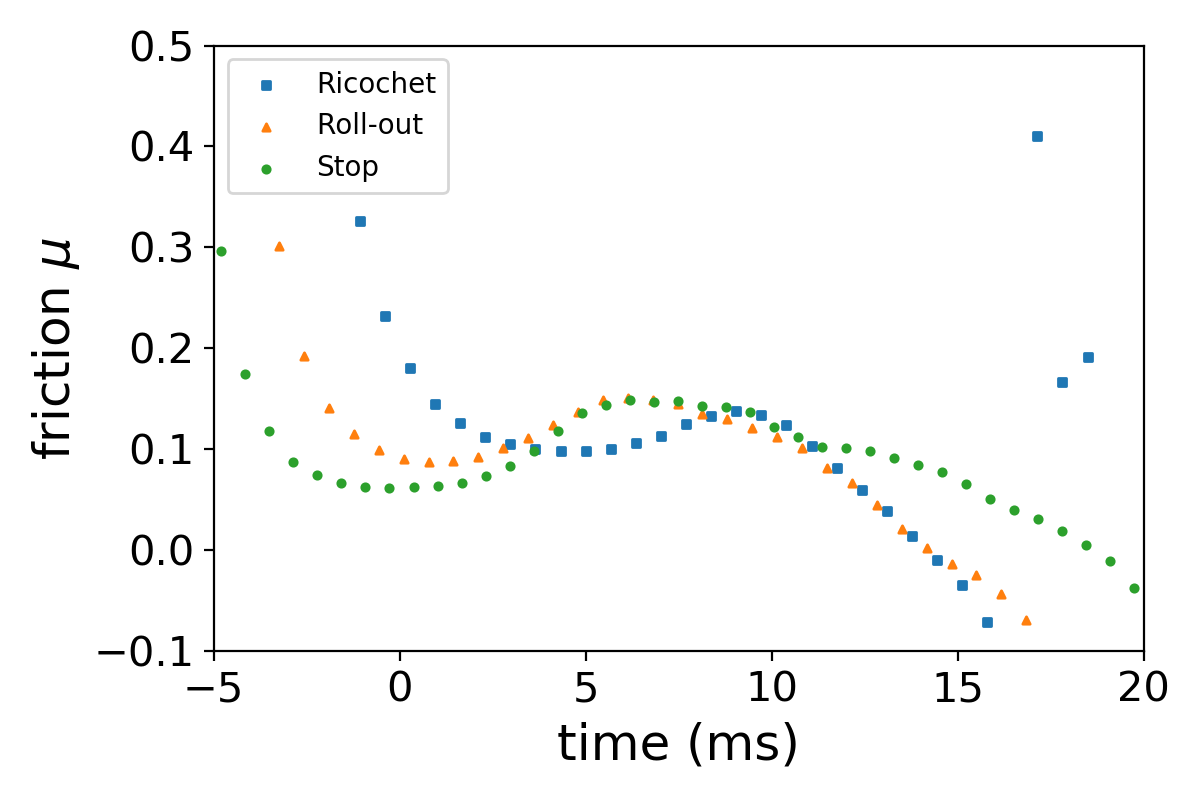}
\caption{The effective friction coefficient between the projectile and the sand for each experiment presented as a function of time from impact. The friction coefficient was computed  using Equation \ref{mu_fric} and  accelerations measured from the tracked positions of the projectile and angular accelerations from the fitted spin data.
\label{fig:friction}
}
\end{figure}

\subsection{Estimating an effective friction coefficient from the spin}

We estimate an effective friction coefficient between marble and sand using the rate of change of spin during the impact.
We assume that the friction force $F_\mu  = \mu_{eff} F_N$ on the marble is set by an effective friction coefficient times a normal force.  We estimate the normal force from the size of the vertical acceleration $F_N = m_p a_z$.  The torque on the marble $\tau = I \dot \omega \approx R_p F_\mu$.   For a homogeneous sphere the moment of inertia $I = \frac{2}{5} m_p R_p^2$. Putting these together, we estimate the friction coefficient 
\begin{equation}
 \mu_{eff} \sim \frac{I \dot\omega}{R_p m_p a_z} \sim \frac{2 R_p \dot \omega}{5 a_z} .
\label{mu_fric}
\end{equation}
Using the rate of change of spin $\dot \omega(t)$ and vertical acceleration $a_z(t)$ from our projectile trajectories
we use equation \ref{mu_fric} to estimate the effective friction coefficient as a function of  time $\mu_{eff}(t)$.
Equation \ref{mu_fric} requires dividing by acceleration. If the acceleration is low, the result is noisy. To mitigate this effect, we only measured the effective friction coefficient during the early and 
high acceleration phase of impact. 

The effective friction coefficient measured near the time of impact for each experiment is shown in Figure \ref{fig:friction}.
We computed the angular acceleration by passing our 
angular rotation rates through a Savinsky-Golay filer with width of 25 samples.
The high values prior to impact are spurious and due to dividing by low value for the acceleration prior to impact.  We only plotted the estimated friction coefficient during the early and high acceleration phase of impact for the same reason.

The coefficient of dynamic friction of glass to glass contact is 0.4. Our measured values peak around 0.1 which is below that expected for sand and marble friction contacts.  Equation \ref{mu_fric} can also be approximated as $\mu_{eff} \approx 2 R_p \Delta \omega / (5 \Delta v_z)$, using a change in spin $\Delta \omega$ and a change  in the vertical velocity component $\Delta v_z$.  Using changes in both quantities during the high acceleration phases of the impacts, we confirm that the estimated friction coefficient is approximately 0.1. This check ensures that our estimate is not affected by how we smoothed the data.  

We found that the value of the  effective  friction coefficient $\mu_{eff} \sim 0.1 $ is remarkably low.
The low value for the effective friction coefficient suggests that sand particles are acting like lubricant, or ball bearings rolling under the marble.  Alternatively,  contacts on the front size of the marble could be partly cancelling the torque exerted from contacts on the bottom of the marble. 

\subsection{Trends}
\label{sec:trends}

Phenomenological models of low velocity impacts into granular media have primarily been developed for normal impacts
(e.g., \citealt{tsimring05,katsuragi07,goldman08,katsuragi13,brzinski13,murdoch17}).
%Empirical force laws for the acceleration  of the projectile during impact are functions of velocity and depth.
The developed empirical force laws are based upon measurements of impact penetration depth, duration and trajectories as a function of depth or time during the impact (e.g., \citealt{katsuragi13}).  To help pin down the force laws
for non-normal impacts we use our trajectories to search for relations between acceleration
and other parameters such as velocity and depth.

\begin{figure*}
\centering
\begin{tabular}{ll}
\includegraphics[width=3.0in,trim = 20 20 0 0, clip]{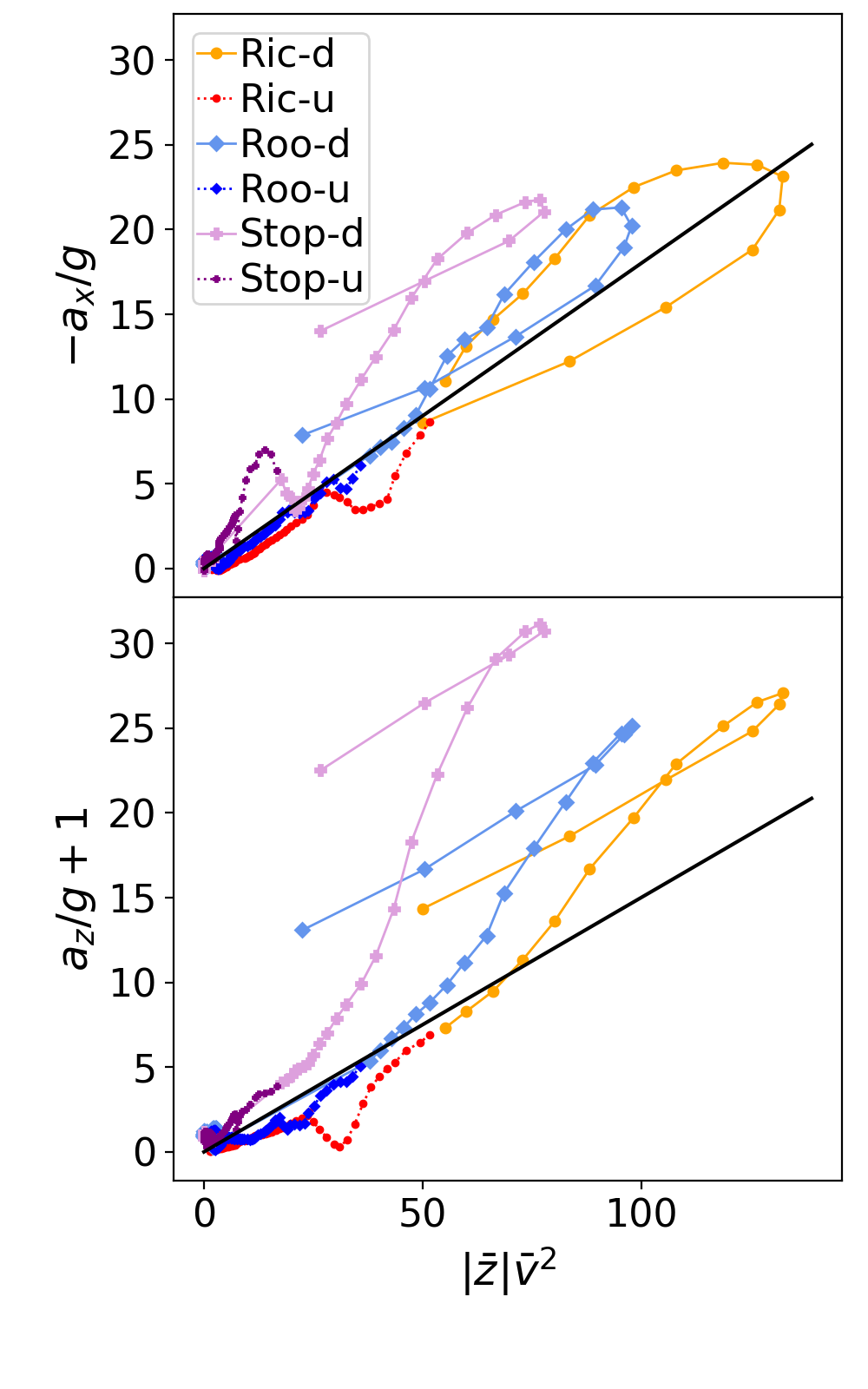}
%\hspace{1cm} 
&
\includegraphics[width=3.0in,trim = 20 20 0 0, clip]{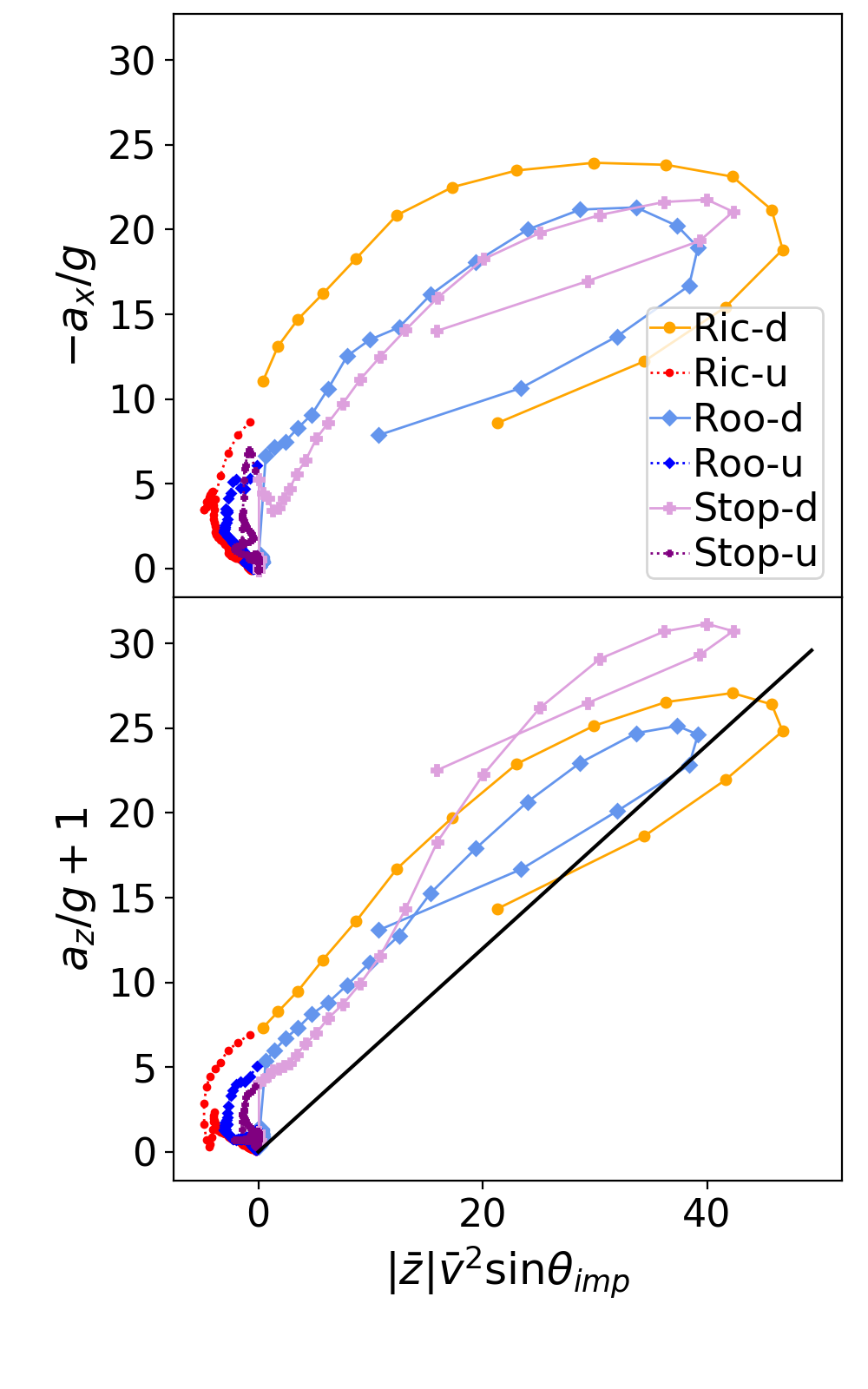}
\end{tabular}
\caption{Accelerations measured from three experiments as a function
of $|\bar z| \bar v^2$ on the left and $|\bar z| \bar v^2 \sin \theta_{impact}$ on the right. Each point is at a different time, and are taken from the trajectory plots shown in Figure \ref{fig:pend_traj}.  The larger and lighter points are during the penetration phase and labeled Ric-d, Roo-d and Stop-d, corresponding to filmed ricochet, roll-out and stop experiments.
The smaller and darker points are during the rebound phase and are labeled Ric-u, Roo-u and Stop-u.
\textbf{Left:} The black fit lines in the upper and lower panels have slopes of 0.18 and 0.15 respectively. These slopes fit the upward rebound phase well but not the early penetration phase.
\textbf{Right:} The black fit line in the lower panel has a slope of 0.6.
This functional form better fits the vertical acceleration during the downward penetration phase.
%The slopes are a decent match to the horizontal and vertical drag forces during the rebound phase of the impacts.
%The $|\bar z| \bar v^2 $ scaling does not match the accelerations during the earlier penetration (downward) phases of these trajectories. 
\label{fig:a_zv2}
}
\end{figure*}

In Figure \ref{fig:a_zv2} 
we show the projectile horizontal and vertical components of acceleration from three experiments and in units of $g$. The accelerations are plotted as a function of combinations of depth $|\bar z| = |z|/R_p$ normalized by the marble radius and velocity in units of $\sqrt{g R_p} = 28.13$ cm/s.
The dimensionless $\bar v  = v/\sqrt{gR_p}$ is akin to a Froude number. Each trajectory is labeled with different colors and marker types. Each point is at a different time with positions, velocity and acceleration shown on our trajectory plots (Figures \ref{fig:pend_traj}). 
The color and marker size depends upon whether the vertical velocity component is positive or negative.  The lighter colors and larger markers are for the initial penetration phases where the projectile is moving downward into the granular media.   The darker colors and smaller markers show the later part of the trajectories when the projectile is moving upward and the velocities are lower.
The colors and point types are shown in the legends with labels ending with 'd' corresponding to the initial downward penetration phase.   Labels  ending with 'u' correspond to the later rebound phase when the projectiles move upward.

In the left panel in Figure \ref{fig:a_zv2} we plot acceleration components as a function 
of $|\bar z| \bar v^2$. We also show gray lines with a slope of 0.18 and 0.15 in the horizontal and vertical directions respectively.  These slopes were found to approximately match both the horizontal and vertical accelerations in the slower and later (upward) rebound phases of the trajectories (and as seen on the lower left side of both panels). In this later phase, the accelerations scale with the square of the velocity, as would be expected from hydrodynamic drag or lift forces.
When the projectile is moving upward, the vertical acceleration should be called `lift' rather than drag as both velocity and acceleration are in the upward direction.  
The $|\bar z| \bar v^2 $ scaling does not match the accelerations during the earlier penetration phases
of these trajectories. 

The right panel of Figure \ref{fig:a_zv2} shows the horizontal and vertical components of the projectile's acceleration as a function of $|\bar z| \bar v^2 \sin \theta_{impact}$. 
%The horizontal component $a_x$ does not match this fit. 
The vertical component $a_z$ has a black fit line with a slope of 0.6.
It can be seen from both panels that the penetration phase of $a_z$ does not scale as $\bar z \bar v^2$ alone but with the sine of the impact angle.

During the penetration phase a drag-like horizontal force with acceleration $a_x \propto -v^2$ is approximately supported by the top left panel in Figure \ref{fig:a_zv2}.  An
upward drag-like vertical force dependent on the square of the velocity and sine of the impact angle seems   
approximately supported by the lower right panel in Figure \ref{fig:a_zv2}.  We will use these relations to leverage prior normal impact studies to approximate estimate the maximum penetration depth and horizontal component of velocity at the time of maximum penetration in section \ref{sec:models}.

%These rough scaling relations suggest that we can use empirical relations based on prior studies of normal impact experiments to estimate the depth and time of maximum penetration and the horizontal  component of velocity at the same time.

We searched for combinations using scaling polynomials of powers of $\bar z$ and components of $\bar {\bf v}$ that would put all the acceleration points on a single curve.  If this were possible, such a curve would have allowed us to create an empirical force law that could be integrated to predict projectile trajectories.   Unfortunately we failed to find simple combinations of depth and velocity for collapsing our trajectories to a single curve.  

Figure \ref{fig:a_zv2} 
illustrates that a single force law does not fit both penetration and rebound phases of the trajectories. The trends seen here suggest that during the rebound phase, the vertical and horizontal forces scale with depth and velocity, and are similar in their dependence.  However the forces during the penetration phase, before the time of maximum depth, must differ in form compared to those in the rebound phase. This might due to compaction of granular medium in front of the projectile, acting like ramp in front of a snow plow  (e.g., \cite{percier11}). Simulations that take into account the response of the granular medium are probably required to match our projectile trajectories.

\section{Phenomenological models}
\label{sec:models}

\subsection{Empirical models for Normal Impacts}

Phenomenological models have been proposed to account for experimental measurements of penetration depth of non-spinning spherical projectiles impacting granular materials in a gravitational field and at normal incidence (e.g., \citealt{ambroso05,tsimring05,katsuragi07,goldman08,katsuragi13,altshuler14,murdoch17}).
The equation of motion of the projectile's vertical position during the impact,
\begin{equation}
   \frac{d^2 z}{dt^2} = -g + \frac{F_d}{m} \label{eqn:motion}
\end{equation}
where $z$ is the vertical coordinate with $z=0$ at the point of impact on the granular surface, $m$ is the projectile mass and $g$ is the downward vertical acceleration due to  gravity.  An empirical form for the vertical force from the granular substrate decelerating the projectile
\begin{equation}
   F_d =  F_z(z) + B(z) v + \alpha v^2 \label{eqn:normal}
\end{equation}
where $F_z(z)$ is a depth dependent force and called a hydrostatic, frictional or quasi-static resistance force term.  The $v^2$ term describes an inertial or hydrodynamic-like drag force (e.g., \citealt{allen57,tsimring05,katsuragi07,goldman08,pachecovazquez11,murdoch17}). Sometimes a velocity dependent term, here with coefficient $B(z)$, is  included that looks like a low Reynolds number drag term (e.g., \citealt{allen57,goldman08}).

The force law of equation \ref{eqn:normal} is commonly only applied while the projectile decelerates (e.g., \citealt{goldman08,katsuragi13}).   
A maximum penetration depth $d_{mp}$, is reached when the vertical velocity first reaches zero.  The time at which this happens (after impact) is called a stopping or collision time $t_s$.   The equation of motion may not have a fixed point at this depth and at this time, so would give subsequent upward acceleration.  However the post penetration phase upward motion is irrelevant when estimating a stopping time 
and a maximum penetration depth.   

Recent normal impacts experiments into granular media using Atwood  machines find that past a certain impact velocity, the maximum penetration depth  is approximately independent of effective gravity and is approximately proportional to the collision velocity, $d_{mp} \propto v_{impact}$  \citep{goldman08,murdoch17}.  
% \citep{katsuragi13,tsimring05}.  % exponents where 2/3, 4/5, but  not Atwood
% debruyn also found exponent of 1
The experiments by \citet{goldman08} have impact velocities of a few m/s and effective gravity 0.1 to 1 g and those by  \citet{murdoch17} have velocities 1 to 40 cm/s and effective gravity 10$^{-2}$ to 1 g.
The collision time or duration is approximately independent of impact velocity and  the effective gravity \citep{goldman08,murdoch17}.
These findings build upon prior experiments at 1 g that found similar scaling laws \citep{debruyn04,ambroso05,ciamarra04,tsimring05,katsuragi13}. 
The equations of motion with empirical force law in the form of Equation  \ref{eqn:normal}
can match the trends measured for the 
maximum penetration depth and collision time, (e.g., \citealt{tsimring05,goldman08,katsuragi13}).  

%Accelerations we saw in our trajectories during the penetration phase of impact, do scale with velocity and depth (as shown in Figure \ref{fig:a_zv2}), so the force law in equation \ref{eqn:normal} might be modified to depend on impact angle and take into account both horizontal and vertical motions.  

% However, we did not see as clear trends in acceleration versus depth and velocity in the rebound phase of the impacts.  We have attempted to model full trajectories with modified force laws and have had difficulty matching them during throughout the trajectory all the way to the final static equilibrium state.

\begin{figure}
\centering
    \includegraphics[width=3.5in]{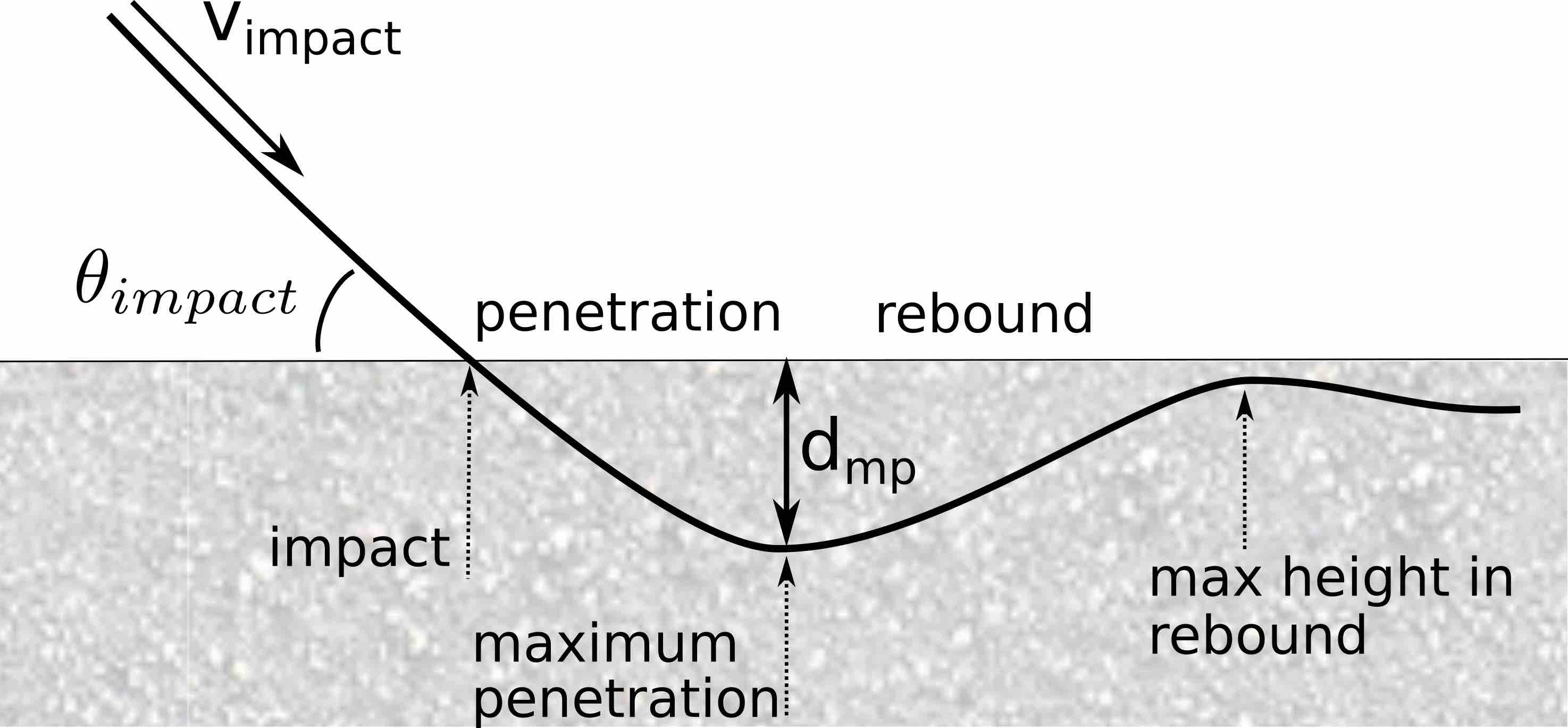}
\caption{Different phases of an impact.
\label{fig:phases}}
\end{figure}

%\subsection{Ricochets}

%Projectiles hitting sand or water at high velocities and low impact angles can skip or ricochet off the surface.   Empirical force laws have also been used to model ricochets  at low impact angles into water and sand \citep{birkhoff44,johnson75,daneshi77,soliman76,bai81}. For spherical projectiles into water or sand, models for a non-spinning sphere   give   a velocity dependent grazing critical impact angle, $\theta_{cr}(v_{impact})$, below which ricochet occurs \citep{birkhoff44,johnson75,daneshi77,soliman76,bai81}.  

%We adopt $\theta_{impact} =0$ for a grazing impact and $\theta_{impact} = 90^\circ$ for a normal impact.

\subsection{Empirical model for ricochet and roll-out lines}

Prior empirical models for ricochet that we  introduced in section \ref{sec:crit} \citep{birkhoff44,johnson75,daneshi77,soliman76,bai81}, assume that the horizontal velocity component is nearly constant and that the lift force depends on it with $F_L \propto  v_x^2$.  The assumed lift is dependent on depth  and computed by integrating a hydrostatic pressure applied to the submerged surface of the spherical projectile.  This pressure is estimated following Bernoulli's principle.  The velocity dependence of the lift force resembles that of the hydrodynamic drag-like force that has been used to model normal impacts (e.g., \citealt{katsuragi07}).   

Our trajectories and trends in them discussed in section \ref{sec:trends} imply that a single empirical model, based on that developed for normal impacts, would not give a good description for both the penetration phase and rebound phases of impacts  (see Figure \ref{fig:phases} for an illustration of these phases).  We attempt to improve upon prior ricochet models in sand by using scaling developed for normal impacts into granular media to estimate a maximum penetration depth.  We then use a simple but different model for the post penetration or rebound phase of impact to estimate a criterion for ricochet and roll-out events.
 
%We attempt to  modify the lift based model for ricochet in sand  to see if a similar model can also predict roll-outs.
%Instead of trying to model both penetration and rebound phases, we focus on  the rebound phase alone.  

At the moment of maximum penetration,  the horizontal velocity is $v_{xmp}$,  the depth is $d_{mp}$ and the vertical velocity component $v_z =0$.  We first estimate $v_{xmp}$ and $d_{mp}$ from the impact angle  $\theta_{impact}$ and the impact velocity $v_{impact}$.
We then find the height reached in the rebound phase from a vertical equation of motion that has a lift force that is dependent on the horizontal component of velocity.

To estimate the horizontal component of velocity at the moment of maximum penetration, we assume
that the horizontal velocity component during the penetration phase is described with a hydrodynamic-like drag 
\begin{equation}
   \frac{dv_x}{dt} = - \alpha_x v_x^2 . \label{eqn:avx}
\end{equation}
Here $\alpha_x$ is a drag coefficient that has units of inverse length. 
For hydrodynamic drag on a sphere of radius $R_p$, the drag force is proportional to the projectile cross sectional area and the drag coefficient
depends on the density ratio and projectile radius, 
\begin{equation}
  \alpha_x \approx \frac{\rho_s}{\rho_p} \frac{3}{4 R_p}. \label{eqn:alphax}
\end{equation}
Equation \ref{eqn:avx} has solution 
\begin{equation}
 v_x(t) = \frac{v_{x0}}{v_{x0} \alpha_x t + 1}  ,\label{eqn:vxt}
\end{equation} 
where the initial horizontal velocity $v_{x0} = v_{impact} \cos \theta_{impact}$.
The horizontal velocity component at the time of maximum penetration 
\begin{equation}
 v_{xmp} 
 %=  \frac{v_{x0}}{v_{x0} \alpha_x t_s + 1}   
 =  
 \frac{v_{impact}  \cos \theta_{impact}}{v_{impact}  \cos \theta_{impact} \alpha_x t_s + 1},  \label{eqn:vxmp}
\end{equation} 
in terms of the stopping time $t_s$.  Following experimental studies of normal impacts \citep{murdoch17}, we assume that the stopping time $t_s$ is independent of velocity and effective gravity. We also assume that the stopping time $t_s$ is independent of impact angle, as supported by recent experiments of oblique impacts \citep{bester19}.

Following experimental studies of normal impacts (e.g., \citealt{goldman08,murdoch17}), we assume that the depth of maximum penetration $d_{mp}$ is proportional to impact velocity.
We expect that the depth of maximum penetration $d_{mp}$ would be lower for shallower impact angles, so we assume
\begin{equation}
d_{mp} = k_d t_s  v_{impact}  \sin \theta_{impact}, \label{eqn:dmp}
\end{equation} 
with unit-less parameter $k_d$. We estimate $\alpha_x$ and $k_d$ from measurements of our tracked trajectories.
This angular dependence is consistent with penetration depth proportional to the initial z component of velocity, and this is approximately supported by the recent oblique impact experiments by \citet{bester19} (see their Figure 2).

We assume that the lift during the rebound phase is proportional to the square of the horizontal velocity component, as did prior models for ricochet \citep{soliman76,bai81}. These ricochet models assumed that the horizontal component of velocity was nearly constant during the impact and that the lift was depth dependent.  Our trajectories show that the horizontal component of velocity varies significantly during rebound. Likely the lift during rebound is both depth dependent and time dependent through its sensitivity to the horizontal velocity component. 

To roughly characterize a regime for ricochet and roll-out we ignore the depth dependence of the lift but we take into account its time dependence. For the rebound phase we use an equation of motion in the vertical direction
\begin{equation}
\frac{dv_z}{dt} = c_L v_x(t)^2 - g \label{eqn:vz_rebound}
\end{equation}
with lift coefficient $c_L$ that is in units of inverse length.
The left term is lift and the right term is the %downward 
gravitational acceleration. With  $z=0$, the bottom edge of the projectile touches the substrate surface and with $z=-2R_p$, it is entirely submerged.

The horizontal component of velocity $v_x(t)$ in the rebound phase follows equation \ref{eqn:vxt} but with a drag coefficient
$\beta_{x}$ that might differ from that present during the penetration phase (that we called $\alpha_x$).
In the rebound phase 
\begin{equation}
 v_x(t) =  \frac{v_{xmp}}{v_{xmp} \beta_{x} t + 1}   \label{eqn:vx_rebound}
\end{equation} 
where $v_{xmp}$ is the horizontal velocity component at the moment of maximum depth (estimated in equation \ref{eqn:vxmp}) and time is measured from the beginning of the rebound phase.
Initial conditions are $v_x(0) = v_{xmp}$, $v_z(0) = 0$ and $z(0) = - d_{mp}$.
We integrate Equation \ref{eqn:vz_rebound} using  Equation \ref{eqn:vx_rebound} for $v_x(t)$
\begin{align}
 v_z(t) &= - \frac{c_L}{\beta_x} \frac{v_{xmp} }{ (v_{xp} \beta_x t + 1)} - gt + \frac{c_L}{\beta_x} v_{xp} \nonumber \\
 & = \frac{c_L v_{xmp}^2 t}{ (v_{xmp} \beta_x t + 1)} - gt.  \label{eqn:vzt}
\end{align}
The constant of integration is determined by requiring $v_z(0) = 0$ at the moment of maximum depth. The maximum height (or minimum depth) during rebound is subsequently reached when $v_z(t_m)=0$ where
\begin{equation}
t_m =\frac{c_L v_{xmp}^2 - g}{v_{xmp} \beta _x}. \label{eqn:tm}
\end{equation}
We integrate Equation \ref{eqn:vzt} to find the height in the rebound phase 
\begin{align}
 z(t) = - \frac{c_L}{\beta_x^2}   \ln (v_{xp} \beta_x t + 1) - \frac{g t^2}{2}+ \frac{c_L}{\beta_x} v_{xp} t - d_{mp}.
 \label{eqn:zt}
\end{align}
The height $z(t_m)$ gives a maximum height during the rebound phase.
 
Let's examine the time $t_m$ (equation \ref{eqn:tm}) which is the time in the rebound phase when height $z(t)$ reaches an extremum.
We require $t_m >0$ for the rebound trajectory to rise and not sink. This gives  condition 
\begin{equation}
  c_L v_{xmp}^2 >g.
\end{equation}
If the horizontal velocity component does not significantly vary during the impact then is equivalent to 
\begin{equation}
  \cos^2 \theta_{impact}  > \frac{g}{v_{impact}^2  c_L}.
\end{equation}
In the limit of low impact angle  this condition becomes
\begin{equation}
 \theta_{impact}^2 <  1 -  \frac{g}{v_{impact}^2 c_L} 
\end{equation}
which is similar to the expression for the critical angle giving ricochet by \citet{soliman76}.
A comparison between this equation and equation \ref{eqn:orange}, the orange
line we adjusted to match the ricochet line on Figure \ref{fig:ric_flip2}, gives
a lift coefficient $c_L \approx 0.02 R_p^{-1}$.   
Henceforth we allow the horizontal velocity component to decay during the impact.
With horizontal drag, a larger lift coefficient would be required for ricochet to occur.

We can use $z(t_m)$ computed using equations  \ref{eqn:vxmp}, \ref{eqn:dmp},  \ref{eqn:tm},  and \ref{eqn:zt} to estimate the height reached during the rebound phase.  These are functions of initial $v_{impact}, \theta_{impact}$,  and coefficients  $\beta_x, c_L,\alpha_x, t_s, k_p$.   The coefficients $\alpha_x, t_s, k_p$ can be estimated from our impact trajectories.  The drag and lift coefficients  $\beta_x, c_L$ can be adjusted.   The result is an estimate of the height $z(t_m)$ as a function of impact velocity $v_{impact}$ and angle  $\theta_{impact}$.

If $z(t_m) >0$ then the projectile rises above the level of the substrate and we would classify the event as a ricochet.
If $0>z(t_m) > -R_p$ then the center of mass of the projectile rises above the substrate level and the projectile could roll.   We assign the condition $z(t_m) = 0$ to be the line dividing
ricochet from roll-out and $z(t_m) = -R_p$ to be the line dividing roll-out from stop events.  By computing  $z(t_m)$ and adjusting $c_L, \beta_x$ to match our experimental event classifications, we find an empirical model for these two dividing lines.

We measured stopping times, $t_s$, maximum penetration depths $d_{mp}$ and horizontal velocity components $v_{xmp}$ at the maximum depth for the three videos that we tracked.
These quantities are listed in Table \ref{tab:max_pen}. 
The three tracked videos have stopping time $t_s \sim 0.01$ s.  
% $\tau_s \approx 0.35$.
From the maximum penetration depths and using equation \ref{eqn:dmp} we estimate the factor $k_p \sim 0.5$.   The horizontal velocity component measured at maximum penetration depth divided by the initial horizontal velocity component is about 0.3 for our three tracked videos.  From the horizontal velocity components measured at the maximum depth, impact angles and velocities and using equation \ref{eqn:vxmp} we estimate the drag coefficient during penetration phase $\alpha_x \sim 0.4 R_p^{-1}$.  This is similar
to that expected for ballistic drag (as estimated in equation \ref{eqn:alphax}) and is consistent with trends seen in the tracked trajectories,
(shown in Figure \ref{fig:a_zv2}).

\begin{figure}
\centering
  \includegraphics[width=3.5in, trim = 10 0 10 0,clip]{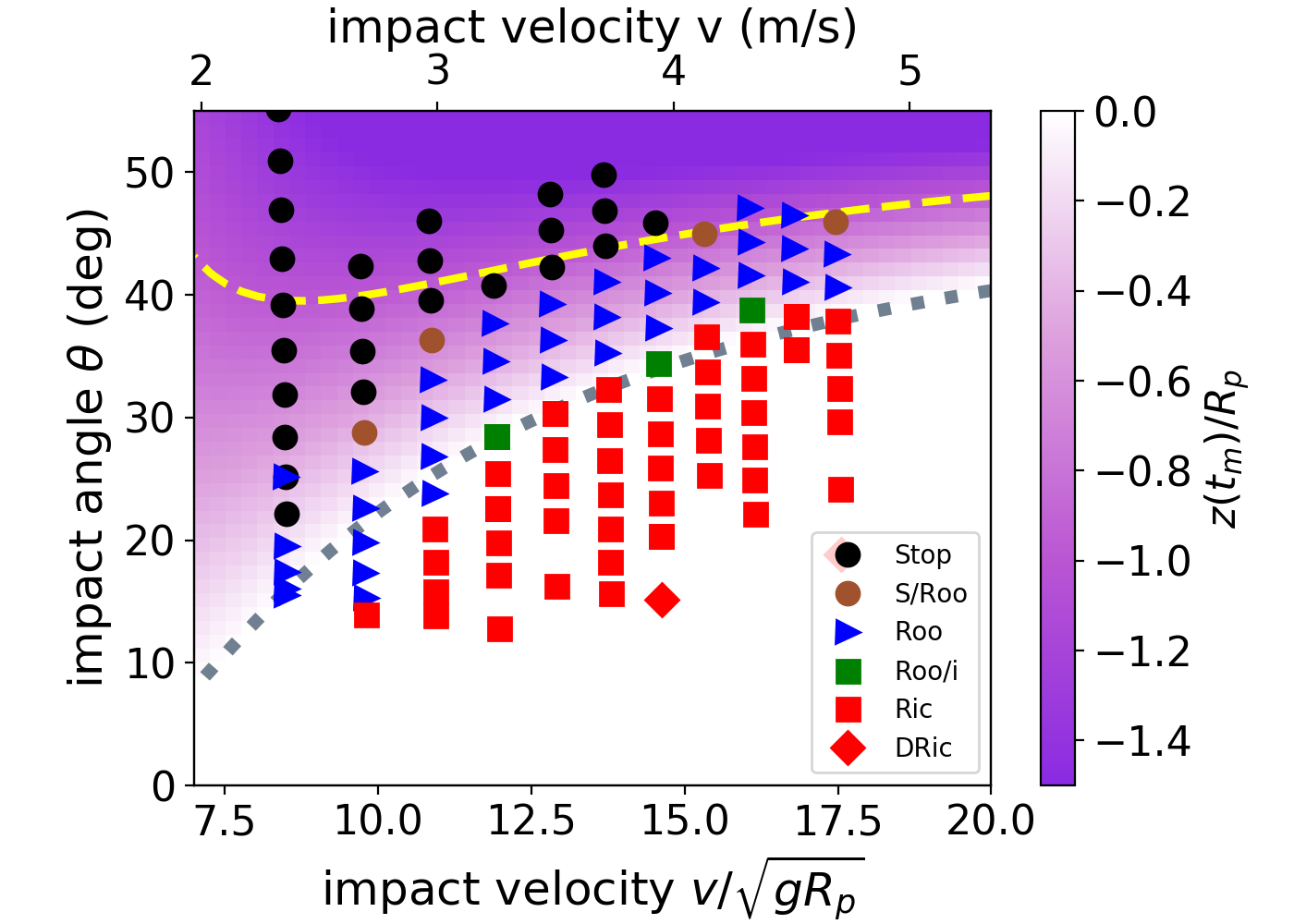}
\caption{
The points show the same experiment events previously shown in Figure \ref{fig:ric} and \ref{fig:ric_flip2}.  Here the  lower $x$ axis is the impact Froude number or the impact velocity in units of $\sqrt{g R_p}$ where $R_p$ is the marble radius.  The top axis gives impact velocity in $m/s$.  The $y$ axis is the grazing impact angle in degrees. The colormap shows the height reached during the rebound phase predicted using  \ref{eqn:vxmp}, \ref{eqn:dmp},  \ref{eqn:tm},  and \ref{eqn:zt}. The height reached is given by the colorbar on the right in units of projectile radius $R_p$. The dotted grey contour line corresponds to empirical model giving  height in the rebound phase of $z(t_m) = 0$.  This is a critical angle for ricochet as the projectile can rise above the substrate during the rebound phase. The yellow dashed line gives $z(t_m) = -R_p$ and is a critical angle allowing the projectile center of mass to rise  above the substrate surface.  This line is an estimate for 
the division between roll-out and stop events.
\label{fig:ric_flip3}}
\end{figure}

In Figure \ref{fig:ric_flip3} we show with a color map the rebound height $z(t_m)/R_p$ computed at different values of impact velocity and angle.  The colorbar shows the value of $z(t_m)/R_p$. The $x$ axis is a Froude number or impact velocity in units of $\sqrt{g R_p}$. The rebound height in the rebound phase was computed using equations  \ref{eqn:vxmp}, \ref{eqn:dmp},  \ref{eqn:tm},  and \ref{eqn:zt} and the above estimated values for stopping time, penetration depth parameter $k_p$ and drag coefficient $\alpha_x$.
The remaining parameters used are lift coefficient $c_L = 0.15/R_p$ and rebound phase drag coefficient $\beta_x = 0.1/R_p$.
On this plot, we also show our experiment impact classifications that were described previously in section \ref{sec:delin} and shown in Figures \ref{fig:ric} and \ref{fig:ric_flip2}.  
The upper dashed yellow line shows a $z(t_m) = -R_p$ contour and the lower white dotted line shows a depth $z(t_m) = 0$ contour.  These are estimates for the critical angle giving roll-out and that giving ricochet.  The model is a pretty good match to the experimental ricochet/roll-out line,  but overestimates  the critical angle for the roll-out/stop line, particularly at lower velocities. The rolling marble, as seen in the roll-out event trajectory shown in Figure \ref{fig:pend_traj}), stays at a particular equilibrium depth while rolling.   A better prediction for the roll-out/stop dividing line might be made by computing the height that lets lower surface of the marble rise above this equilibrium level during the rebound phase.

We find that rebound drag coefficient must be smaller than the penetration phase drag coefficient, $\beta_x < \alpha_x$.  Otherwise, the ricochet line on Figure \ref{fig:ric_flip3} does not rise with increasing velocity. A lower value of the rebound drag coefficient is consistent with the shallow slope in $v_x$ seen in Figure \ref{fig:pend_traj} in the rebound phases.
There is some degeneracy between rebound phase drag and lift coefficients, $\beta_x$ and $c_L$.  This degeneracy is not surprising, since their ratio appears in equation \ref{eqn:zt} for the height reached during rebound.  Extremely low values of $c_L$ would give rebound phases that are longer than we observed.

As was true for prior penetration depth and stopping time estimates (e.g., \cite{katsuragi13}), our model does not have an equilibrium fixed point at the maximum height reached during the rebound phase.   After this height is reached, the gravitational acceleration in the model would cause the projectile to drop forever.  A more complete model could add a hydrostatic-like force term, dominating at low velocity, that allows the projectile to reach a final equilibrium resting condition at a shallow depth.  We opted to use a time dependent but depth independent lift force and a constant gravitational acceleration. The result is an acceleration that is approximately linearly dependent on time.  It might be possible to 
derive a similar looking model with depth dependent forces.  We attempted to do so with constant but depth dependent hydrostatic and lift terms but had less success with them.

% \begin{table}  %nomen
% \centering
% \caption{Nomenclature \label{tab:nomen}}
% \begin{tabular}{lll}
% \hline
% Projectile mass & $m_p$ \\
% Surface gravitational acceleration & $g$ \\
% Projectile radius, if spherical & $R_p$ \\
% Granular substrate mean density & $\rho_s$ \\
% Projectile density & $\rho_p$ \\
% Projectile velocity at impact  & $v_{impact}$ \\
% Projectile velocity vector & ${\bf v}$ \\
% Projectile cross sectional area & $A$ \\
% Critical impact angle & $\theta_{cr}$ \\
% Froude number  & Fr $= \bar v = v/\sqrt{R_pg}$ \\
% Horizontal coordinate & $x$  \\
% Vertical coordinate & $z$ \\
% Depth below surface level  &  $|z|$ with $z<0$ \\
% Impact angle & $\theta_{impact}$ \\
% %Spherical coordinate angles & $\phi,\psi$ \\
% Drag force & $F_d$ \\
% Lift force & $L$ \\
% Coefficient of static friction  & $\mu_s$ \\
% Angle of repose  & $\theta_r$ \\
% Stopping time & $t_s$ \\
% Maximum penetration depth & $d_{mp}$ \\
% Horizontal velocity component  & $v_{xmp}$ \\
% ~ at maximum depth & \\
% Time of maximum height & $t_m$ \\
% ~ during rebound & \\
% Height reached in rebound & $z(t_m)$ \\
% Drag coefficients & $\alpha_x, \beta_x $ \\
% Lift coefficient & $c_L$ \\
% \hline
% \end{tabular}\\
% Notes.  $\theta=0$ for a grazing impact.  The vertical coordinate is positive upward.
% The horizontal coordinate is positive with the initial direction of projectile motion.
% \end{table}

\section{Application to low-g  environments}

Using dimensionless numbers and scaling arguments laboratory experiments can be used to predict phenomena in regimes that are difficult to reach experimentally.
With that idea in mind we discuss using 
scaling laws developed for crater impacts and ejecta curtains \citep{holsapple93} to apply the results of our  laboratory results at 1 g  to asteroid surfaces. 
It is important to note that these scaling laws were developed for hypervelocity impacts normal to the surface, and for  point source impactors (i.e. impactors that are much smaller than the diameter of the crater).
%application of our lab experiments in Earth's gravity $g$ and in air to asteroid environments (vacuum and $10^{-4}g$).
%The scaling relations derived by \citep{holsapple93} for impact craters relate the mass of the ejecta curtain to the projectile mass.

\cite{holsapple93} defines three dimensionless parameters that have historically been denoted  $\pi_2, \pi_3$ and $\pi_4$.  These  dimensionless are used to give regimes and  scaling relations for the crater efficiency, $\pi_1$ (sometimes called $\pi_V$), which is the ratio of the crater mass to the projectile mass.
The first of the dependent dimensionless variables is $\pi_2 \equiv  ga/U^2$ where $a$ is the radius of the projectile and $U$ is its velocity. This is the same as the inverse of the square root of the Froude number. The $\pi_2$ parameter is defined as the ratio of the lithostatic pressure to the dynamic pressure generated by the impact at a depth of the projectile's radius.
The next dimensionless parameter $\pi_3 = Y/\rho U^2$ is the ratio of the crustal material strength $Y$ to the dynamic pressure of the impact $\rho U^2$. The strength of our sand is low and we can assume that regolith on an asteroid will also be low compared to the dynamic pressure from an impact. This results in a small value for $\pi_3$ and so we can neglect it in our scaling argument.
The last parameter, $\pi_4$ is the ratio of the substrate to the projectile density, which in our experiments was 0.64. \citeauthor{holsapple93} ignores this parameter in the scaling relations since its value is confined to be near unity. Since our density ratio is not significantly  different than unity we follow him by neglecting this parameter as well.
This leaves only a single important dimensionless parameter, $\pi_2$ which is directly dependent on the Froude number.

Are there additional dimensionless parameters that might be important in the granular impact setting that were not considered by \cite{holsapple93}? The ratio of projectile radius to grain size radius might be important.  In our experiments this ratio is large (and equal to about 32).  The ratio is large enough that it probably does not affect our experimental results, however planetary surfaces can have both larger and smaller sized particles near the surface.

We adopt the assumption that we can scale our laboratory  experiments to a low g asteroid environment by matching the Froude number.  Future experimental studies at low g facilities and using granular media of different size distributions could test this assumption.

% Ejecta curtains on asteroids may be more massive due to the milli and micro gravity regime compared to the ejecta we see in our experiments. \textbf{Also, if ejecta is more massive then make snow plow argument?}

We estimate the conditions (the velocity and impact angle) that would allow a rock to ricochet on asteroids such as Bennu or Ryugu. The escape velocity from a spherical object of radius $R_a$ can be written in terms of its surface gravity $g_a = GM/R_a^2$
\begin{equation}
v_{esc} = \sqrt{\frac{2GM}{R_a}} = \sqrt{2 g_a R_a}.
\end{equation}

We can write impact velocity in units of $\sqrt{g_a R_p}$ as
\begin{equation}
 \bar v^2 =  \frac{v_{impact}^2}{{g_a R_p} }= 2 {\frac{R_a}{R_p} } \left (\frac{v_{impact}}{v_{esc}}\right)^2.
 \label{eqn:va}
\end{equation}
We insert this velocity into equation \ref{eqn:orange} for the critical 
angle allowing ricochet, giving us the critical angle
as a function of projectile and asteroid radius.  These are plotted on Figure \ref{fig:regime}. The series of black, red, and orange lines are for impacts at the escape velocity.  Each line is labelled with the critical impact angle and points to the right and below the line allow ricochets below this labelled impact angle.  The series of blue and green lines are for ricochets at a tenth of the escape velocity.
The axes on this plot are log10 of asteroid and projectile diameters in meters.

Figure \ref{fig:regime} shows that  few meter diameter and smaller boulders on 500 m diameter asteroid  such as Bennu, when hitting a region of level granular material at the escape velocity, would be likely to ricochet.  Since most impacts would not be normal, this would apply to a large fraction of such objects.  At lower velocities  ricochets would only be likely for few cm sized objects.

We find that large boulders, such as the 14 m one shown in Figure \ref{fig:bennu} would be above the ricochet line and so would not ricochet.  However this size boulder is near enough to the ricochet line that it might roll upon impact.  If use equation \ref{eqn:gray}
instead of equation \ref{eqn:orange} to make this plot, then a 14 m diameter rock is on the line allowing roll-out to take place at impact angles below $30^\circ$.

If Froude number is relevant for matching grazing impact behavior at 1 g to that on asteroids, then we infer that boulders on Bennu would have rolled or ricocheted upon impact for near escape velocity impacts. 

\subsection{Ricochets on Eros}

% track 1,2,3 boulders are 40, 2m, 25 m long, from fig 1 caption durda+12
% estimated speeds are 7, 3, 5 m/s, respectively from Tables 1,2,3 
% elevation angles are 10-30 degrees
% surface gravities are 4.4e-3, 2.7e-3 and 4.2e-3 (page 1095) 
%r = 20.; g = 4.4e-3; v = 7.0; print v/sqrt(g*r)
%r = 1.; g = 2.7e-3; v = 3.0; print v/sqrt(g*r)
%r = 25./2; g = 4.2e-3; v = 7.0; print v/sqrt(g*r)
% Froude numbers are  24, 57,30

\begin{figure}
\centering
  \includegraphics[width=3.5in]{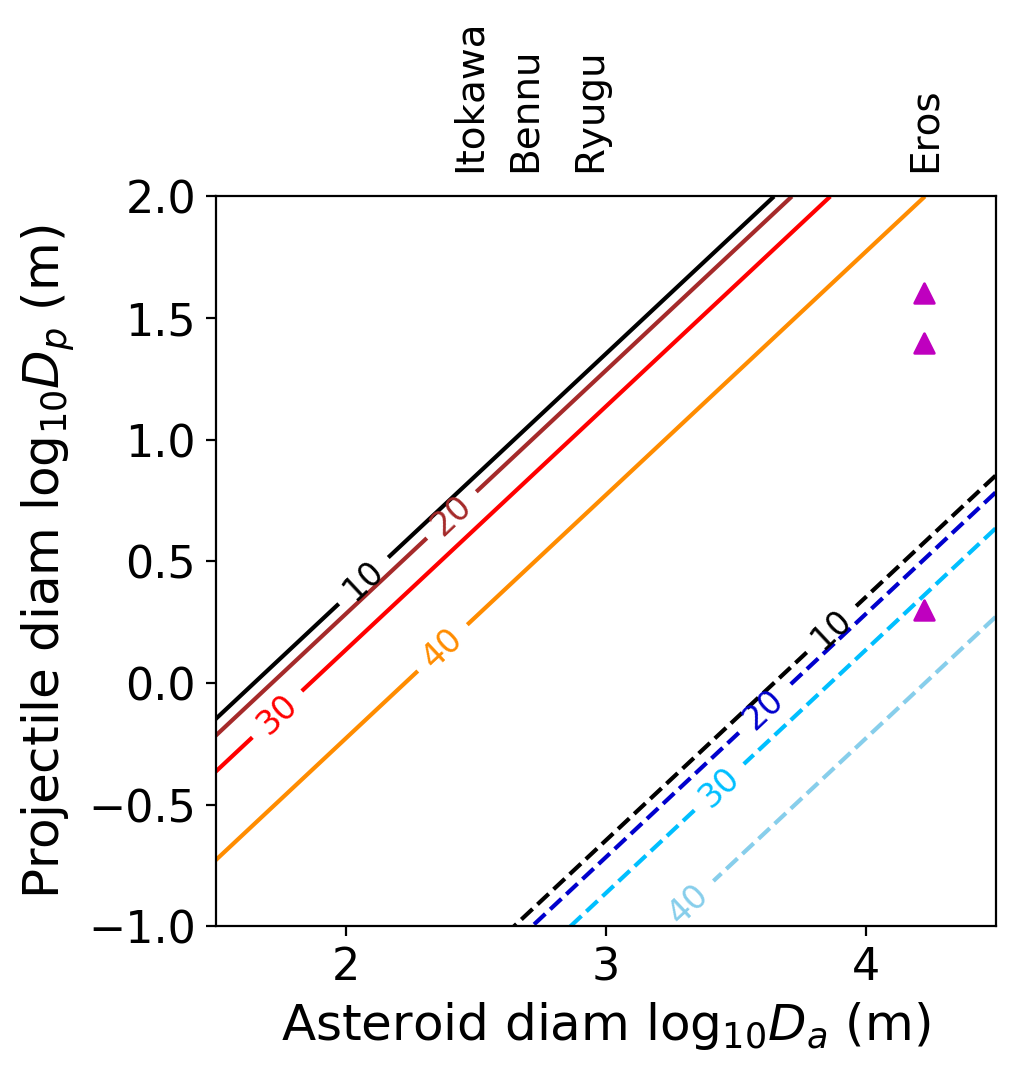}
\caption{
Projectile diameters on different asteroids that would ricochet.  The $x$ axis is asteroid diameter, the $y$ axis is projectile diameter. The lines show critical angles allow ricochet and are computed using the empirical relation of equation \ref{eqn:orange} and \ref{eqn:va}.  The black, brown, red and orange set of solid lines is for impact velocities at the escape velocity.  The black, blue and green dashed set of lines is for impact velocities at 0.1 that of the escape velocity.  The angle allowing ricochet for each line is labeled in degrees. Ricochets occur below and to the right of each line. Smaller and faster objects are more likely to ricochet.
Purple triangles show three boulders on Eros with diameters 40, 25, and 2 m. These were identified with accompanying tracks by \cite{durda12}  who inferred that they impacted with velocities near the escape velocity and ricocheted.  %These lines were computed using Equations \ref{eqn:orange} and \ref{eqn:va}.
\label{fig:regime}}
\end{figure}

\citet{durda12} give examples of 3 tracks or oblong craters, paired with boulders,  that likely made the tracks by skipping of the surface on asteroid 433 Eros. 
The three boulders have sizes 40, 2 and 25 m,
respectively (see their Figure 1).
The estimated impact velocities are $v_{impact} \sim 7, 3, 5$ m/s respectively  (taking typical values from their Tables 1-3). 
Surface gravitational accelerations at the sites of the secondary craters computed
by \citet{durda12} are 
$4.4 \times 10^{-3}$, $2.7\times 10^{-3}$ and $4.2\times 10^{-3}$ m\ s$^{-2}$, respectively.
Using the sizes, velocities and accelerations  values we compute Froude numbers
of 24, 57, 30 where we have used half 
the length scale in place of radius in equation \ref{eqn:Fr}.
They estimate grazing impact angles (which they call the mean elevation
angle) $\theta_{impact} \sim 20^\circ$.  
This places the inferred ricochets on Eros on the lower right hand side of Figure \ref{fig:ric_flip3} and consistent with our estimate for
the division between roll-out and ricochet events. The three boulder sizes observed by \cite{durda12} are plotted as purple triangles on Figure \ref{fig:regime}.
The range of escape speeds on Eros is 3 to 17 m s$^{-1}$  \citep{yeomans00} so these events
are consistent with impact velocity near the escape velocity.  These points lie below the red lines in Figure \ref{fig:regime} and are consistent with the limiting impact angles  velocities for ricochet in our extrapolated model.

The coefficient of restitution of Eros' surface was estimated by \cite{durda12} for their observed tracks. Their estimated values were found by taking the ratio of the projectile's rebound speed and the impact speed with values in the range 0.08-0.19.
Taking the same ratio of our ricochets results in a similar value of 0.14.
These values also closely match our measured effective friction coefficient (see Figure \ref{fig:friction}) during impact.

% \textbf{We get a list of boulders and look up surface accelerations and surface slopes and sizes for them?
% We then tabulate or graph conditions for ricochet for all of them?}

\section{Summary and Discussion}
 
We have carried out laboratory experiments of glass spherical projectiles (marbles) impacting level sand at a range of impact angles. Impact velocities range from 2--5 m/s and grazing impact angles (measured with 0 corresponding to a grazing impact) range from about 10 to 50$^\circ$.   Our projectile material (glass) has density  similar to that of the grains in the granular substrate (sand).  We use a pendulum projectile launcher to reduce projectile spin.

We use high speed camera images to track projectile motion and spin. The projectiles spin up when they penetrate the sand, however, the  friction coefficient required is low, suggesting that the sand particles fluidize near the projectile and effectively lubricate the projectile surface.

We find that projectiles can ricochet or roll-out of their initial impact crater, and that this is likely at higher impact velocities and lower grazing impact angles. This trend is opposite to that found from experiments at higher velocities and higher projectile density into sand that were done by \citet{soliman76}.

We delineate lines between ricochet, roll-out and stop events as a function of impact velocity and angle.  The dividing lines for these classes of events are empiricaly matched by quadratic relation for the square of the critical impact angle, that is in the same form as that derived by \citet{soliman76}, but has larger coefficients. We explore  an empirical model for the post maximum penetration (rebound) phase of  impact, balancing a lift force that is dependent upon the square of the horizontal velocity component against gravitational acceleration. This model estimates a maximum height reached in the rebound phase of the impact. 
A condition for ricochet is the projectile center of mass reaching a maximum height rising above the surface. 
A maximum height just reaching, but not above, the surface gives a condition for projectiles that roll-out of their impact crater.
With adjustment of lift and drag coefficients, 
this empirical model can match our experiment ricochet and roll-out dividing lines. 
 
The projectile trajectories show different scaling in penetration and rebound phases, making it difficult to find simple empirical force laws for the impact dynamics that cover inertial regimes and a lower velocity end phase. Likely a numerical simulation that includes an inertial regime for the granular medium is required to fully understand low velocity or shallow impact dynamics at oblique angles into granular media.  Extending resistive force theory (e.g., \citealt{ding11}) into the inertial and low gravity regimes could be one way to improve empirical models. 
 
We have tried to simplify our experiments by minimizing projectile spin and using spherical projectiles.  In future we would like to explore how spin direction and rate affects the impact. Non-spherical projectiles  would be harder to track  but also interesting to characterize. We would also like to explore the role of irregularities in the substrate, surface level variations and different granular size distribution.
  
Due to their small size, our  projectiles are in a high Froude number regime  where $v/\sqrt{g R_p} \sim 8$ to 17.   If this dimensionless number governs behavior in low gravity environments then our projectiles match m sized projectiles near  the escape velocity  on small asteroids such as Bennu.  The large range of angles allowing ricochet would then imply that projectiles in this regime would predominantly be found distant from their impact crater.  Boulders and accompanying tracks on Eros support this scenario \citep{durda12}.  Experiments in effective low surface gravity could be used to better understand low velocity impact phenomena in low g environments and improve upon our extrapolated models.

\vskip 0.3 truein

\vskip 1 truein %acknow
\textbf{Acknowledgements}

This material is based upon work supported in part by NASA grant 80NSSC17K0771, and National Science Foundation Grant No. PHY-1757062.

We thank Jim Alkins for helpful discussions regarding machining.
We are grateful to Tony Dimino for advise, lent equipment, and suggestions that inspired and significantly improved this manuscript.

\vskip 0.3 truein
 
%{\bf Bibliography}

%\bibliographystyle{mnras}
\bibliographystyle{elsarticle-harv}
\bibliography{ricochet_v04}

% \newpage

% \begin{appendix}
% \section{Appendix}

% In this appendix we describe some additional error checks.

% In Figure \ref{fig:reflection} we show that as the marble traverses the camera field of view, 
% the position of the white light reflection measured 
% with respect to the center of the marble  does not vary significantly.  The angle of the white
% light with respect to the camera angle does not vary enough to cause significant errors
% in our measurements of the marble center of mass, that are based on the position of 
% the white light reflection on the marble surface.

% \begin{figure}
% \centering
% \includegraphics[width=3.5in,trim = 0 60 0 60]{pend030_relfection.png}
% \caption{Panels showing the white light reflection on the marble as a dark black spot in the pend030 video. The frames are shifted such that the center of the marble is at the center of the panels. The left and center panels show the marble at early and late time in the video, respectively. The right panel is the sum of the center and right panels. The reflection does not appreciably move across the marble's surface during the experiment. Variations in illumination angle does not add a significant error to our measured trajectory positions.
% \label{fig:reflection}}
% \end{figure}

% \end{appendix}

\end{document}